\documentclass[twocolumn,10pt]{article}
\usepackage[T1]{fontenc}
\usepackage[utf8]{inputenc}
\usepackage{mathptmx}
\usepackage[scaled=.90]{helvet}
\usepackage{courier}

\usepackage[a4paper,top=0.9in,bottom=1in,left=0.9in,right=0.9in,includefoot]{geometry}
\usepackage{graphicx}
\usepackage{xcolor}
\usepackage{colortbl}
\usepackage{cuted}
\usepackage{microtype}
\usepackage{parskip}
\usepackage{tcolorbox}
\tcbuselibrary{skins,breakable}
\usepackage{academicons}
\usepackage{fontawesome5}
\usepackage{hyperref}
\usepackage{xurl} 
\usepackage{lastpage}
\usepackage{tikz}
\usetikzlibrary{shapes.geometric,arrows.meta,positioning,fit,backgrounds,calc}
\usepackage{booktabs}
\usepackage{tabularx}
\usepackage{caption}
\usepackage{enumitem}
\usepackage{fancyhdr}
\usepackage{titlesec}
\usepackage{float}
\usepackage[ruled,noline]{algorithm2e}
\usepackage{wrapfig}
\usepackage{amsmath, amssymb}
\usepackage{mathtools}
\usepackage{multicol}
\usepackage{multirow}
\usepackage{amsthm}

{\theoremstyle{remark}}
{\theoremstyle{remark}}

\usepackage{cite}
\usepackage{stfloats}

\newcommand{\rowsep}{\arrayrulecolor{black!15}\hline\arrayrulecolor{black}}

\graphicspath{{results/figures/}{figs/}}

\definecolor{primary}{HTML}{003049}
\definecolor{accent}{HTML}{F77F00}
\definecolor{textgray}{HTML}{2F2F2F}
\definecolor{dividergray}{HTML}{DADADA}

\hypersetup{colorlinks=true,linkcolor=primary,citecolor=primary,urlcolor=primary}

\titleformat{\section}
  {\color{primary}\sffamily\Large\bfseries}
  {\thesection}{1em}{}[\vspace{0.2em}\titlerule]
\titleformat{\subsection}
  {\color{primary}\sffamily\large\bfseries}
  {\thesubsection}{1em}{}
\titlespacing*{\section}{0pt}{6pt}{3pt}
\titlespacing*{\subsection}{0pt}{4pt}{2pt}

\setlist[itemize]{left=1.2em, itemsep=2pt, topsep=2pt, parsep=0pt}
\setlist[enumerate]{left=1.2em, itemsep=2pt, topsep=2pt, parsep=0pt}

\tcbset{
  abstractstyle/.style={enhanced,colback=white,colframe=primary,boxrule=1pt,
    fonttitle=\bfseries\sffamily\large,left=6mm,right=6mm,top=2mm,bottom=2mm,
    width=\textwidth,boxsep=4pt,breakable}
}

\makeatletter
\def\@maketitle{%
  \begin{center}
    {\fontsize{18pt}{21pt}\selectfont \bfseries \textcolor{primary}{A Global Readiness and Sovereignty Capability Model for Post-Quantum Cryptography Migration}}\\[1ex]
    {\normalsize
      Mohamed Aly Bouke\,%
      \raisebox{0.6ex}{\href{https://orcid.org/0000-0003-3264-601X}{\includegraphics[height=1.4ex]{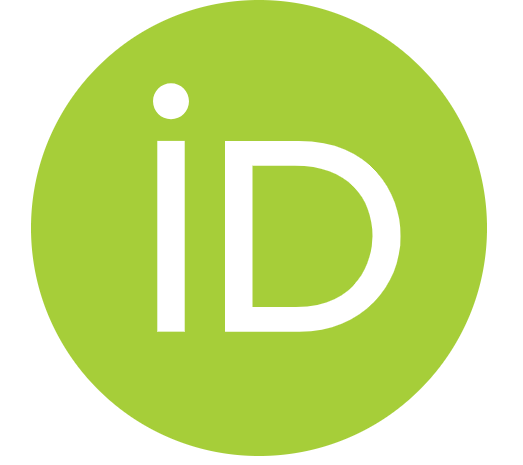}}}%
      \,\raisebox{0.6ex}{\href{mailto:bouke@ieee.org}{\textcolor{gray}{\scriptsize\faEnvelope}}}%
      \raisebox{1.3ex}{\scriptsize1,*}%
    }\\[0.8ex]
    {\footnotesize
      \textsuperscript{1}Centre for Intelligent Cloud Computing, CoE for Advanced Cloud,\par
      Faculty of Information Science and Technology,\par
      Multimedia University, Jalan Ayer Keroh Lama, Bukit Beruang, 75450, Melaka, Malaysia
    }\\[0.5ex]
    {\scriptsize \texttt{*bouke@ieee.org}}\\[1ex]
    {\scriptsize \textit{Research Article}, \today}
  \end{center}
}
\renewcommand{\maketitle}{\twocolumn[\color{textgray}\@maketitle\vspace{-1.2em}]}
\makeatother

\begin{document}
\maketitle

\begin{strip}
  \begin{center}
    \begin{tcolorbox}[abstractstyle, title=Abstract]
      \normalsize
      Cryptographic dependence predates the quantum era, but the migration to post-quantum cryptography
      (PQC) opens a rare window to reshape it, because the algorithms, implementations, hardware, and
      standards adopted now can lock in dependence or sovereignty for decades. This paper
      introduces the Readiness--Sovereignty Capability Model (RSCM), a national measurement model that
      operationalizes PQC readiness together with cryptographic sovereignty, which current maturity models
      score only as readiness and the sovereignty literature defines without measuring. RSCM decomposes sovereignty into three distinct constructs, indigenous cryptographic
      capacity, indigenous post-quantum control, and external dependency, and certifies a
      post-quantum maker only through a gate requiring demonstrated, institutionally sustained
      creation in at least one core layer, whether design, implementation, or validation.
      Applying it to fifty-seven documented cryptographic actors coded from cited public evidence,
      and testing that coding with an independent second coder, a plausible-state bootstrap, and
      convergent-validity checks, we find that twenty countries clear the gate, fifteen as full-stack
      makers and five as research makers, eleven hold strong general capacity without post-quantum
      control, one is a ready adopter, and twenty-five are dependent. The gate cells show substantial
      weighted agreement, a quadratic-weighted kappa of 0.71, and the maker classification is stable in
      its core though uncertain at the threshold. Readiness
      tracks independent cyber indices at rank correlations up to 0.70, while post-quantum creation shows
      no significant correlation with the commitment index, a rank correlation of only 0.22 that
      separates control from readiness. The paper contributes the framework, the evidence-graded
      assessment, and policy directions for building indigenous quantum-safe
      capacity.

      \vspace{0.5em}
      \textbf{Keywords:} post-quantum cryptography; cryptographic sovereignty;
      quantum-safe migration; crypto-agility; digital sovereignty.
    \end{tcolorbox}
  \end{center}
\end{strip}

\section{Introduction}\label{sec:intro}
The standardization of module-lattice and hash-based algorithms by the United States
National Institute of Standards and Technology (NIST) in 2024 turned the migration to
post-quantum cryptography (PQC) from a research concern into a national program
\cite{nist_fips203,nist_fips204,nist_fips205}. The migration is urgent because encrypted
traffic captured today can be stored and decrypted once a cryptographically relevant
quantum computer exists, a threat model known as harvest now and decrypt later. Several
governments have responded with binding timelines and inventory obligations
\cite{nsm10,cnsa2,ombm2302,uae_encryption}.

Assessing national preparedness has so far meant measuring readiness, that is the presence
of a strategy, a cryptographic inventory, standards adoption, deployment, and skills. A
second question receives far less structured attention, namely whether the algorithms,
vendors, and hardware that carry a national migration are controlled domestically or
supplied from abroad. This second question is not new to the quantum era, since a country's
classical cryptographic stack already rests to varying degrees on foreign algorithms, vendors, and
hardware, and cryptographic dependence long predates post-quantum cryptography. What the transition
changes is not the existence of that dependence but the occasion to re-decide it, because a country can
mandate and deploy PQC while depending entirely on foreign standardized algorithms, foreign vendors, and
foreign hardware security modules. This distinction between readiness and sovereignty is invisible to a
one-dimensional readiness score, yet it determines whether a migration produces durable autonomy or only
relocates dependence into the quantum era.

The stakes of getting this distinction right are high and long-lived, and they follow from path
dependence. Cryptography underpins every other digital service, so a dependency at the cryptographic
layer is a dependency of the entire digital country, and the confidentiality lifetimes at risk, which
for government, health, and identity records run to decades, mean that a choice made during this
transition will bind a country well beyond the tenure of those who make it. Because every country is
rebuilding its cryptographic stack at once and against near deadlines, the algorithms, implementations,
hardware, vendors, standards, and skills a country puts in place now are likely to remain the foundation
of its quantum-safe stack for decades, and recovering sovereignty once that foundation has set is slower
and more costly than building it into the migration. The transition is therefore not the origin of
cryptographic sovereignty but a rare window in which countries reshape it for a generation, and
measuring only readiness during that window can credit a rapid but externally supplied migration as
progress while missing the dependence it entails, which is what the second axis is designed to detect.

Three gaps motivate this work. First, published maturity models we reviewed assess the readiness of
products or organizations rather than nations, and do not score sovereignty
\cite{pqcmm,qrmm}. Second, the idea of cryptographic sovereignty is now discussed in policy
and industry writing, but in the sources we found it is defined rhetorically rather than operationalized
into a measured, reproducible national instrument \cite{qnu_sovereignty,navigating_frontier}.
Third, the assessment literature concentrates on a small set of Western and East Asian
leaders, while Africa and the Arab world, whose constraints differ sharply, are covered by
journalism and policy briefs rather than systematic analysis \cite{unu_inequality,gsma_pqc}.

We address these gaps with RSCM, which scores readiness alongside a decomposition of
cryptographic sovereignty into three distinct constructs, general indigenous cryptographic
capacity, indigenous post-quantum control, and external dependency, so that a country's classical
capability is never mistaken for control of the post-quantum layer. Sovereignty over the
post-quantum layer is established through a maker gate that requires demonstrated indigenous
post-quantum technical capability in at least one core layer of design, implementation, or
validation, sustained institutionally rather than as an isolated academic result. RSCM is
operationalized through a coded rubric anchored to cited public evidence, so that scores
are checkable rather than assertive. We position RSCM against product and
organizational maturity models, against the broad and non-reproducible sovereignty indices
that have begun to appear, and against qualitative strategy surveys, and we apply it to a
purposive sample of fifty-seven documented cryptographic actors, with deep coverage of the
under-studied countries of the Global South, presented as a demonstration of the model rather than a
definitive ranking of individual countries.

The paper delivers the following contributions.
\begin{itemize}
  \item RSCM, a national model that scores post-quantum readiness alongside a decomposition of
  cryptographic sovereignty into indigenous capacity, indigenous post-quantum control, and external
  dependency, with a strict post-quantum maker gate that recognizes demonstrated indigenous creation in
  at least one core layer of the quantum-safe stack on an institutionally sustained basis, and a capacity-by-creation typology that
  keeps classical strength from being read as post-quantum sovereignty.
  \item A methodology for defensible, uncertainty-aware scoring, comprising a codebook anchored to
  cited public evidence with an audit trail, decision rules specified in the codebook before coding, per-cell evidence grades
  with a plausible-state bootstrap, an independent second-coder reliability study, robustness checks,
  and convergent validation of readiness against independent indices, alongside construct-alignment
  checks for the sovereignty dimensions.
  \item An evidence-graded assessment of fifty-seven documented cryptographic actors worldwide,
  spanning post-quantum makers, capable but dependent countries, and the under-studied Global South.
  \item A set of policy directions for building indigenous quantum-safe capacity in
  dependent countries, with a focus on the Global South.
\end{itemize}

The remainder of the paper is organized as follows. Section~\ref{sec:related} reviews
related work and positions the model. Section~\ref{sec:background} gives the technical and
policy background. Section~\ref{sec:model} presents RSCM.
Section~\ref{sec:method} describes the methodology. Section~\ref{sec:results} reports the
assessment. Section~\ref{sec:discussion} discusses the findings.
Section~\ref{sec:threats} presents limitations and threats to validity. Section~\ref{sec:recommend} gives
recommendations, and Section~\ref{sec:conclusion} concludes.

\section{Related Work}\label{sec:related}

\subsection{Migration and maturity models}
Several maturity models describe how far an organization or a product has progressed toward
quantum safety. The Post-Quantum Cryptography Maturity Model grades the readiness of products and
services in a supply chain, which suits procurement decisions in which a buyer must judge a
vendor \cite{pqcmm}, and related models grade the internal posture of an organization across
people, process, and technology so that an enterprise can plan its own migration \cite{qrmm}. The
technical migration literature complements these with guidance on crypto-agility and on the
sequencing of discovery, prioritization, and replacement \cite{crypto_agility}. These
instruments are valuable for the tasks they were built for, but three properties limit them for a
national sovereignty assessment. They take the organization or the product as the unit of
analysis rather than the country, so they cannot aggregate to a national position. They measure
progress on a single readiness dimension, so a laggard and a leader differ only in how far each
has advanced along one axis. And they are silent on whether the capability being graded is
domestically controlled, so they cannot express the difference between a nation that authors its
own algorithms and one that adopts them, which is the very difference that decides whether a
national migration builds autonomy or merely relocates dependence. Recent comparative work has moved
beyond organization-level maturity models toward institutional migration analysis. El~Bizri et al.\
compare the transition approaches of major cybersecurity agencies and propose an
institutional--sectoral convergence framework linking policy, standardization, certification, and
risk-management pillars to sectoral readiness \cite{elbizri2026}, and a systematic review of eighteen
published lifecycle models scores them on governance, sequencing, risk, and resource management
\cite{williams2026ssrn}. This work advances comparative migration assessment, but its object remains
migration strategy, lifecycle governance, and sectoral readiness rather than national control of the
underlying cryptographic capability. The present model keeps the ordinal, evidence-anchored spirit of a
maturity model while changing the unit to the country and adding the missing axis, national control of
the post-quantum layer.

\subsection{National strategy and sovereignty}
A second body of work examines national strategy and the politics of the quantum transition, and
it is where the idea of cryptographic sovereignty has taken shape. Policy briefs warn that the
migration can deepen international inequality and create new dependence on foreign vendors,
placing the least-resourced countries at the greatest risk \cite{unu_inequality}. Industry writing
defines quantum-safe sovereignty as the ability of a nation to protect its critical systems with
technology it controls, standards it trusts, and infrastructure it owns \cite{qnu_sovereignty}.
Strategy surveys narrate government roadmaps toward quantum autonomy and situate them in the
wider contest over technological leadership \cite{navigating_frontier}, comparative policy analyses
rank national quantum ambitions and reputations across countries \cite{weber2025dgap}, and broader
indices that mix cyber, artificial-intelligence, and digital-sovereignty capacity into a single
sovereign score have begun to appear \cite{qssi}. This literature establishes convincingly that
sovereignty is the right second question, and it supplies the vocabulary of control, dependence, and
autonomy that the present model formalizes. Most of it treats the concept rhetorically or narratively,
while recent benchmarking work has begun to operationalize related notions of sovereignty \cite{qssi}.
A definition tells a reader what sovereignty is; it does not tell a minister whether a particular
national migration has it, still less how one country compares with another. The broad sovereign
indices, for their part, aggregate so many heterogeneous factors that a single score cannot be traced
back to a checkable cryptographic fact, and they are not reproducible from public evidence. What is
missing is a reproducible cross-national instrument that separates general cryptographic capacity,
post-quantum-specific control, and external dependency, each grounded in cited public evidence.

\subsection{Quantum-safe risk scoring}
A third strand scores technical risk rather than national capability. Quantum-safe risk
assessment frameworks estimate the exposure of an organization or a system to the quantum threat,
weighing the sensitivity and lifetime of data against the maturity of its protection
\cite{quantum_risk_eval}, and resilience scores grade the quantum-safety of concrete computing
and Internet-of-things systems so that an operator can prioritize remediation. This work is
methodologically close to ours in that it too builds an ordinal, criterion-based score, and it is
complementary in substance, since a national capability and a system-level exposure are different
and both matter. But it operates at the level of an asset or an organization and answers how
exposed a system is, not how autonomous a nation is, and it says nothing about who controls the
cryptography that a system depends on. A system can be fully patched with the standardized
algorithms and score as low-risk while the nation that operates it has no control over those
algorithms at all, which is exactly the gap between exposure and sovereignty that the present
model is built to measure.

\subsection{Digital sovereignty and supply chains}
The present work also connects to a broader literature on digital sovereignty and on dependence in
critical technology supply chains \cite{digital_sovereignty}. That literature argues that control
over foundational digital technologies, from semiconductors to cloud infrastructure to
cryptography, has become a dimension of national power, and that a country dependent on foreign
suppliers for a foundational technology inherits both a security risk and a loss of policy
autonomy. Cryptography is a particularly acute case within this literature, because it is the
technology on which the confidentiality and integrity of every other digital service ultimately
rest, so dependence at the cryptographic layer is not confined to one service but propagates
through the entire digital order. The quantum transition sharpens the case further, because it
forces every country to rebuild its cryptographic supply chain at once and on a deadline. The
contribution of the present model to this literature is to make the abstract concern concrete for
one foundational technology, replacing a general argument about dependence with a measured,
country-by-country account of who controls the cryptography and who does not.

\subsection{Synthesis and positioning}
The four strands leave a specific opening. Maturity models and the recent comparative migration
frameworks measure readiness, lifecycle governance, or sectoral convergence but not national control of
the cryptographic capability, and most are not country-level in unit. The strategy and sovereignty
literature raises sovereignty but largely does not measure it. Risk-scoring frameworks measure exposure
at the system level. To our knowledge, we found no existing instrument that jointly measures national
post-quantum readiness together with a decomposed, evidence-traceable account of cryptographic
sovereignty, separating general cryptographic capacity, post-quantum-specific creation and control, and
external dependency, and gating the maker label on demonstrated, traceable capability. RSCM occupies this
opening. It is national, it decomposes sovereignty into capacity, post-quantum control, and
external dependency, it gates post-quantum control on demonstrated capability, and it is built on a
coded rubric with an audit trail, which together turn sovereignty from a slogan into a measurement.
Table~\ref{tab:position} sets the model against the closest existing instruments along the
dimensions that matter for a national sovereignty assessment.

\begin{table*}[t]
\centering\footnotesize
\setlength{\tabcolsep}{4pt}
\renewcommand{\arraystretch}{1.2}
\caption{RSCM against the closest instruments along seven dimensions. It alone is national in unit,
measures sovereignty, separates indigenous from adopted capability, cites evidence for every score,
and is comparable across countries.}
\label{tab:position}
\begin{tabularx}{\textwidth}{@{}>{\raggedright\arraybackslash}p{2.3cm} >{\raggedright\arraybackslash}p{1.9cm} >{\raggedright\arraybackslash}p{1.9cm} >{\centering\arraybackslash}p{1.5cm} >{\centering\arraybackslash}p{1.6cm} >{\centering\arraybackslash}p{1.5cm} >{\centering\arraybackslash}X@{}}
\toprule
Instrument & Unit & Sovereignty & Indig.\ vs adopted & Evidence traceable & Cross-national & Repro. \\
\midrule
Maturity models \cite{pqcmm,qrmm} & Product or organization & Not measured & No & Self-assessed & No & Partly \\ \rowsep
Sovereign indices \cite{qssi} & Country, broad & Asserted & No & No & Broad & No \\ \rowsep
Risk scores \cite{quantum_risk_eval} & System & Not measured & No & Technical & No & Varies \\ \rowsep
Sovereignty concept \cite{qnu_sovereignty} & Nation, concept & Defined only & No & No & No & No \\ \rowsep
Strategy surveys \cite{navigating_frontier} & National strategy & Narrated & No & Qualitative & Partly & No \\ \rowsep
\textbf{RSCM} & \textbf{Nation} & \textbf{Decomposed} & \textbf{Yes} & \textbf{Yes, cited} & \textbf{Yes} & \textbf{Computational} \\
\bottomrule
\end{tabularx}
\end{table*}

\section{Background}\label{sec:background}

\subsection{The harvest-now quantum threat}
A cryptographically relevant quantum computer would run Shor's algorithm at a scale sufficient to
compromise the widely deployed public-key schemes based on integer factorization and discrete
logarithms, so that the key-establishment and signature mechanisms built on them would fail together
rather than one at a time. Estimates of when such a
machine will exist vary widely, from about a decade to considerably longer, and that uncertainty is
often read as a reason to wait. The harvest-now-and-decrypt-later threat removes that comfort.
Because an adversary can capture and store ciphertext today and decrypt it once the machine exists,
the effective deadline for any given data item is not the arrival of a quantum computer but that
arrival minus the length of time the item must remain confidential.

This arithmetic moves the deadline into the present for exactly the data a country most needs to
protect. Diplomatic and intelligence records, health and genomic data, and identity and financial
records carry confidentiality lifetimes measured in decades, so even a distant quantum computer
places their effective protection deadline in the near term, and for the longest-lived secrets it
has passed already. The transition must therefore begin well before a quantum computer is
demonstrated, which is the reasoning behind the binding national deadlines that several governments
have set ahead of any demonstrated capability, and it is why a wait-and-see posture is itself a
decision to expose long-lived data. The urgency is also unevenly distributed, since a country holding
large archives of long-lived secrets faces a nearer effective deadline than one that does not,
which is part of what the readiness axis is built to capture.

\subsection{Standardized algorithms and hybrids}
The standardized replacements are a module-lattice key-encapsulation mechanism, a
module-lattice digital signature algorithm, and a stateless hash-based signature scheme
\cite{nist_fips203,nist_fips204,nist_fips205}. The first two rest on the presumed hardness of
structured lattice problems, which are believed to resist both classical and quantum attack,
while the third rests only on the security of a hash function and serves as a conservative
fallback whose assumptions are the most thoroughly understood. The concentration of the two
primary schemes on a single mathematical family is itself a source of systemic risk, because a
future advance against structured lattices would affect key establishment and signing at once,
and it is one reason several national agencies require the standardized schemes to be deployed
in hybrid combination with a classical algorithm or an additional conservative post-quantum
scheme, so that a weakness in any single new algorithm does not by itself expose the traffic
\cite{bsi_pqc,anssi_pqc}. Migration is complicated further by the scale of embedded
cryptography, which makes discovery and inventory a prerequisite for any transition, by the
performance and message-size differences of the new schemes, which stress constrained protocols
and devices, and by the long tail of protocols, libraries, and hardware security modules that
must be updated in a coordinated sequence rather than all at once. A distinct response to the same
threat should be set aside at the outset, since quantum key distribution establishes keys through a
physical channel rather than by replacing an algorithm, requires specialized optical hardware, is
limited in range and topology, and protects key exchange rather than the broad landscape of
signatures and authenticated protocols the standardized algorithms cover. Several countries in this
study operate quantum key distribution pilots while having no post-quantum algorithm deployment, so
treating the two as equivalent would credit a narrow physical link as a national migration; the
assessment therefore never scores a quantum key distribution deployment as post-quantum deployment,
a rule applied uniformly alongside the separation of announced plans from operational capability.

\subsection{The national mandate landscape}
National responses vary in force, and the variation is the empirical ground on which the
readiness axis is built. The United States couples binding deadlines for national security
systems with annual inventory obligations, and it sets the migration horizon that many other
countries echo \cite{nsm10,cnsa2,ombm2302}. The European Union frames the transition around
technological sovereignty and prepares supporting legislation \cite{eu_quantum}, while national
agencies in Germany and France issue their own migration profiles and hybrid requirements that
diverge from the pure adoption of the international standard \cite{bsi_pqc,anssi_pqc}. The
United Kingdom, Canada, Australia, Singapore, and India each publish national timelines that
track the leading agencies while adopting the standardized algorithms
\cite{ncsc_uk,ccs_canada,asd_ism,csa_sg,india_nqm}. China pursues a parallel track built on its
own commercial cryptography suite and a national post-quantum call \cite{china_sm,china_iccs},
which raises the prospect that the world will migrate to more than one algorithm family rather
than a single global suite. In the Arab region the United Arab Emirates has issued a binding
national encryption policy that requires transition plans and cryptographic discovery
\cite{uae_encryption}, the only binding mandate in the regional core. Most other countries, by
contrast, have general cybersecurity strategies with no post-quantum content, which is itself a
finding rather than a gap in coverage.

\subsection{Crypto-agility}
Because the standardized algorithms may themselves be revised, and because divergent national
suites may coexist, the durable objective of a migration is not the one-time installation of a
particular algorithm but crypto-agility, the ability to replace cryptographic primitives
repeatedly and at low cost \cite{crypto_agility}. Crypto-agility depends on an accurate
cryptographic inventory, on abstractions that separate applications from the primitives they
call, and on governance that can mandate a change across an estate. It also depends, at the
national level, on the capacity to detect that a change is needed and to obtain a trusted
replacement, which is precisely where sovereignty enters, since an agile country that must still
source every replacement from abroad has automated its dependence rather than removed it. The
transition is therefore best read as the acquisition of a standing national capability, and the
model measures how much of that capability each country controls.

\subsection{Standardization and its geography}
The algorithms now being deployed emerged from a multi-year public process in which candidate
schemes were submitted, analyzed, attacked, and winnowed before a small number were standardized
\cite{nist_fips203,nist_fips204,nist_fips205}. The process was open in that anyone could submit
and analyze a candidate, and its openness is the main reason the resulting standards are trusted,
but the authority to select and to standardize rested with a single national institute, and the
winning designs, while internationally authored, are promulgated as that institute's standards.
The geography of authorship therefore differs from the geography of standardization, and both
differ from the geography of adoption. A European academic team may design a scheme, a United
States agency may standardize it, and a ministry in the Global South may adopt it, and each of
these is a different relationship to the technology. The model's algorithmic and governance
dimensions are built to capture this separation, crediting authorship and standard-setting where
they occur rather than assuming that adoption of a good algorithm confers control over it. The
parallel Korean and Chinese processes \cite{kpqc,china_iccs,china_sm} show that the selection
authority need not be singular, and they raise the prospect that the coming decade will see more
than one standardized family in use, which makes the position of an authoring or standard-setting
country materially different from that of a pure adopter.

\subsection{The economics of the transition}
The transition is expensive and its costs fall unevenly, which is part of why sovereignty and
readiness diverge. Discovering and cataloging the cryptography embedded across a national estate,
re-engineering protocols and applications to accommodate larger keys and signatures, replacing or
reconfiguring hardware security modules, and retraining staff together constitute a multi-year
program whose cost scales with the size and age of a country's digital infrastructure. A wealthy
country can fund this program and reach high readiness quickly by purchasing foreign products and
services, which is precisely the path that produces high readiness with low sovereignty, since buying
it can often be the faster and lower-cost short-term route to a completed migration. Building
indigenous capacity, by contrast, tends to be slower and more expensive in the short run and pays off
only over the longer horizon across which primitives will be replaced repeatedly. The economic logic therefore pushes almost
every country toward adoption, and the external-dependency construct measures the long-run cost of that
short-run choice, so that a country can weigh the cost of dependence against the cost of capacity rather
than only the latter.

\section{The RSCM Framework}\label{sec:model}
\begin{figure*}[t]
  \centering
  \includegraphics[width=0.82\textwidth]{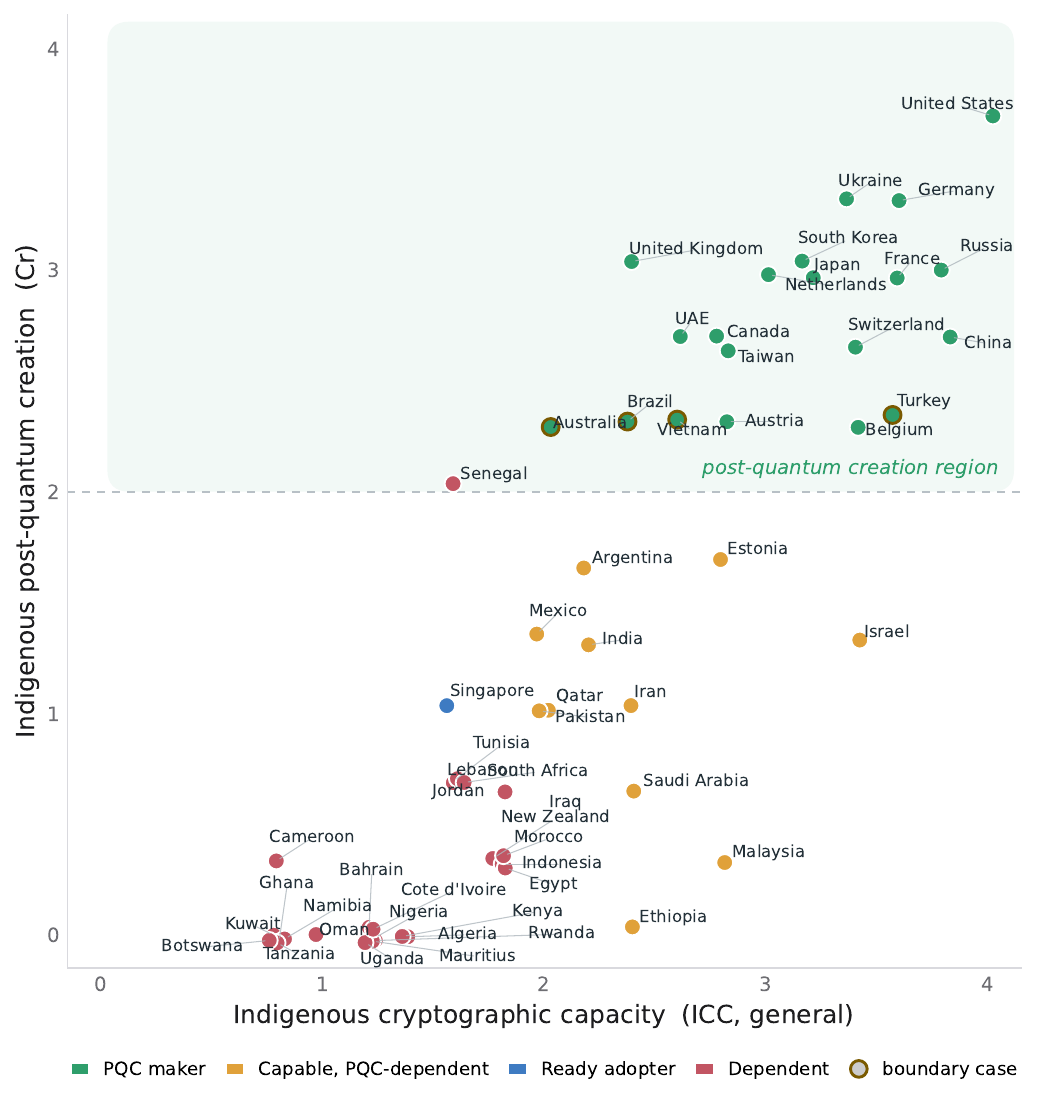}
  \caption{The capacity-by-creation plane, indigenous cryptographic capacity ICC (horizontal) against
  indigenous post-quantum creation Cr (vertical). The shaded band marks high creation, not the gate;
  makers sit in it, dependents at the origin, and boundary cases are ringed.}\label{fig:plane}
\end{figure*}

RSCM scores each country on readiness and on a decomposed sovereignty. Readiness measures how far a
national migration has progressed. Sovereignty is not one axis but three, because a single sovereignty
score conflates a country's general cryptographic capacity, its control of the post-quantum layer
specifically, and its external dependency, and these are held separately. Readiness and indigenous
cryptographic capacity are each the mean of five ordinal dimensions scored zero to four against the
codebook of Table~\ref{tab:codebook}, in which every level is tied to a required public artifact;
indigenous post-quantum control is read from five objective sub-variables; and a country is a
post-quantum maker only when it clears a gate on the post-quantum layer, so classical or institutional
strength alone never earns the label. Figure~\ref{fig:plane} places the countries on the plane of
capacity against post-quantum creation that the typology uses.

\subsection{The readiness dimensions}
Readiness has five dimensions, ordered roughly from intention to execution, with the full level
anchors given in Table~\ref{tab:cb_r}. Strategy and mandate
captures whether a country has taken a national position at all, and at the top level whether that
position is a binding legal instrument with an enforced deadline rather than an aspirational
document, because a mandate with teeth is what actually moves an estate. Inventory captures
whether the country can discover and catalog its cryptography, which is the prerequisite that every
migration guide places first, since an organization cannot replace what it cannot see. Standards
adoption captures whether the standardized algorithms or national profiles have been taken up, in
guidance, in pilots, or in a binding profile. Deployment captures movement from awareness through
pilots to production in critical sectors, and it excludes quantum key distribution, which is a
different response to the same threat. Workforce captures the training and skills base, from ad
hoc awareness to a sustained national pipeline, because a migration that outruns the people who
can execute it stalls. The five are scored separately rather than collapsed, so that a country with
a strong strategy but no deployment is distinguished from one with quiet production and no
strategy.

\subsection{Defining cryptographic sovereignty}
Cryptographic sovereignty over the post-quantum transition is the degree to which a country controls
the quantum-safe cryptography on which its digital order will depend, meaning the capacity to design,
implement, validate, govern, and if necessary repair or replace the post-quantum primitives,
implementations, and hardware it deploys. Three facts about the construct organize the rest of the
model. First, sovereignty over the transition is not the same as general cryptographic capacity: a
country may operate a mature national public-key infrastructure, a domestic vendor base, and an
indigenous classical algorithm, and still control none of the post-quantum layer it is migrating to.
Second, sovereignty is not autarky: a country that adopts a widely scrutinized foreign standard such
as the standardized module-lattice mechanism may retain substantial sovereignty if it controls the
implementation, validation, deployment, key management, and replacement of that primitive, so using a
foreign standard is not by itself a loss of control. Third, control is layered and the layers can be
held separately, so a single sovereignty number conflates distinct capabilities. The model therefore
decomposes sovereignty into three distinct constructs rather than one, and reserves the label of
post-quantum maker for demonstrated indigenous creation in at least one core layer of the post-quantum
stack.

\subsection{Decomposing sovereignty}
The first construct is indigenous cryptographic capacity (ICC), the general and largely
classical-inclusive ability to produce and control cryptography. It is the mean of five layers scored zero to four from
cited evidence against the codebook of Table~\ref{tab:cb_a}: algorithmic research and authorship, the vendor and supply chain, national
infrastructure operation, governance and standards authority, and the research and talent base. A country
with a national trusted-algorithm project, a domestic vendor base, and a national public-key
infrastructure scores high here whether or not any of it is post-quantum, which is the correct
reading of a real national capability rather than a claim about the quantum transition.

The second construct is indigenous post-quantum control, the distinctive quantity of the paper, and
it credits only the post-quantum layer. It is read from five objective sub-variables coded separately
rather than from a single judgment, against the codebook of Table~\ref{tab:cb_b}: whether the country has an indigenous post-quantum primitive, a
nationally standardized or state-backed quantum-safe algorithm as distinct from an individual academic
co-authorship; a domestic post-quantum implementation it can build and maintain; an independent
post-quantum validation and evaluation capability; control of national post-quantum integration and
migration; and the agility to modify or replace a deployed primitive on its own, coded as a standing
capacity rather than a demonstrated swap. Coding the layers
separately, rather than deriving a single indigenous, adopted, or mixed tag, keeps the most
consequential judgment diagnosable and addresses the low reliability that a single control tag was
found to carry.

The third construct is external dependency (ED), coded zero to four, recording how much of the
post-quantum stack a country deploys but does not control, from a self-sufficient national stack to a
fully foreign one. A capable adopter is expected to score high here, and this dependency, restricted
to the components a country deploys without controlling, is the borrowed post-quantum sovereignty
that a one-dimensional sovereignty score hides.

\subsection{The post-quantum maker gate}
A country is a post-quantum maker only if it demonstrates indigenous post-quantum-specific capability
in at least one core creation layer, primitive design, implementation, or validation, and that
capability is institutionally sustained rather than an isolated academic activity. The gate is
deliberately strict on two points where a one-dimensional sovereignty score fails. Classical or
institutional strength alone never qualifies, so a national classical algorithm, a governance seat,
or a national public-key infrastructure is credited as general capacity but not as post-quantum
control, and a country whose only indigenous cryptography is classical and whose post-quantum
primitives are adopted does not become a maker on that basis. And algorithm authorship is treated as
strong evidence rather than the sole requirement, so a country that does not originate a primitive but
can implement, harden, validate, and replace post-quantum cryptography domestically still clears the
gate through those layers. Writing $b_1$, $b_2$, and $b_3$ for the ordinal levels, each coded from
zero to four, of the three creation layers, primitive design, implementation, and validation, and
$\mathcal{I}$ for the set of countries whose only qualifying creation cell rests on isolated,
non-sustained academic activity, a country $c$ is a post-quantum maker exactly when
\begin{equation}
  \mathrm{maker}(c)\;\Longleftrightarrow\;\max(b_1,b_2,b_3)\ge 3 \;\wedge\; c \notin \mathcal{I}, \label{eq:gate}
\end{equation}
where $\max(b_1,b_2,b_3)$ is the strongest of the three creation layers, $\wedge$ is logical
conjunction, and $\mathrm{maker}(c)$ is the resulting maker verdict, true or false.

Requiring a single core layer at level three, rather than several, is deliberate and is the reason
the maker label is always paired with a type. The gate marks one boundary, the line between a country
that creates nothing in the post-quantum layer and one that has demonstrably created in at least one
of design, implementation, or validation, and crossing that line is a real change in kind that a
country either has or has not made. It is not a claim of broad control across all creation layers; the full-stack maker type specifically
denotes demonstrated design and implementation, its stricter condition being level three on both, and
the gate asserts no more than that for any other type. A research maker that has designed a standardized primitive but builds
nothing, and an implementation maker that builds schemes it did not originate, are
each genuinely sovereign in one layer and dependent in the others, and the type records exactly which,
so a single strong layer earns the maker label without being mistaken for control of the whole stack.
Within the makers, the creation profile names one of four types. A full-stack maker both designs and
implements post-quantum cryptography ($b_1\ge 3$ and $b_2\ge 3$); a research maker designs it without a
domestic production base ($b_1\ge 3$, $b_2<3$); an implementation maker implements it without
originating a primitive ($b_1<3$, $b_2\ge 3$); and a validation maker clears the gate through
validation alone ($b_1<3$, $b_2<3$, so $b_3\ge 3$). The distinction separates a country that authored a
standardized primitive from one that can build post-quantum cryptography it did not design, different
kinds of sovereignty that a single maker label would equate. The type is a qualifying threshold rather than a depth ranking,
recording which creation layers clear the level-three anchor rather than how deep or broad the national
base is, so two countries of the same type can differ widely in scale. Depth within a type is read from
the ordinal scores, the indigenous capacity ICC and the creation score Cr, and from the plane of
Figure~\ref{fig:plane}, not from the type label.

\begin{table*}[t]
\centering\small
\caption{Condensed codebook. Each dimension is scored zero to four against cited public evidence.}
\label{tab:codebook}
\begin{tabularx}{\textwidth}{@{}l l X@{}}
\toprule
Construct & Dimension & Level-4 anchor \\
\midrule
Readiness & R1 Strategy and mandate & binding legal instrument with a named enforced deadline and critical-sector scope \\ \rowsep
Readiness & R2 Inventory & mandated continuous cryptographic inventory across critical infrastructure \\ \rowsep
Readiness & R3 Standards adoption & mandated national post-quantum profiles across government and critical systems \\ \rowsep
Readiness & R4 Deployment & broad production across critical sectors (quantum key distribution excluded) \\ \rowsep
Readiness & R5 Workforce & sustained pipeline of degree tracks, national certification, and active groups \\ \rowsep
Capacity & A1 Algorithmic authorship & nationally standardized indigenous algorithm, classical or post-quantum \\ \rowsep
Capacity & A2 Vendor and supply chain & domestic supply chain including primitives and hardware \\ \rowsep
Capacity & A3 Infrastructure operation & sovereign national stack including domestic hardware \\ \rowsep
Capacity & A4 Governance and standards & sets standards adopted by others or authors an international standard \\ \rowsep
Capacity & A5 Research and talent base & sustained national research ecosystem producing talent and IP \\ \rowsep
PQC control & B1 Indigenous PQC primitive & nationally standardized indigenous post-quantum algorithm \\ \rowsep
PQC control & B2 PQC implementation & full domestic post-quantum implementation and toolchain \\ \rowsep
PQC control & B3 PQC validation & authoritative national post-quantum evaluation program \\ \rowsep
PQC control & B4 National PQC integration & full sovereign control of national post-quantum integration \\ \rowsep
PQC control & B5 PQC replaceability & agility plus authorship or standardization to design a replacement \\
\bottomrule
\end{tabularx}
\end{table*}

\begin{table*}[t]
\centering\footnotesize
\caption{Readiness codebook. Each dimension is scored on the ordinal level whose indicator the cited
evidence meets.}
\label{tab:cb_r}
\begin{tabularx}{\textwidth}{@{}l XXXXX@{}}
\toprule
Dim & 0 & 1 & 2 & 3 & 4 \\
\midrule
R1 strategy and mandate & no public position & awareness statements only & announced or draft strategy, no deadline & adopted strategy with a sector deadline & binding legal instrument with an enforced deadline \\ \rowsep
R2 inventory & none & guidance recommends inventory & pilot inventory in some agencies & national discovery program or tool available & mandated continuous inventory across critical infrastructure \\ \rowsep
R3 standards adoption & none & standards referenced in guidance & pilots using standardized post-quantum schemes & government profiles issued & mandated national profiles \\ \rowsep
R4 deployment & none & awareness only & pilots or proofs of concept & production in some critical sectors & broad production across critical sectors \\ \rowsep
R5 workforce & none & ad hoc training or awareness & university courses or some professionals & national skills program or agency teams & sustained pipeline of tracks and certification \\
\bottomrule
\end{tabularx}
\end{table*}

\begin{table*}[t]
\centering\footnotesize
\caption{Indigenous cryptographic capacity codebook, the general and classical-inclusive base.}
\label{tab:cb_a}
\begin{tabularx}{\textwidth}{@{}l XXXXX@{}}
\toprule
Dim & 0 & 1 & 2 & 3 & 4 \\
\midrule
A1 algorithmic authorship & foreign only, no research & general crypto research & active research groups & indigenous construction or national competition & nationally standardized indigenous algorithm \\ \rowsep
A2 vendor and supply chain & foreign vendors only & foreign vendors with local integrators & local products on foreign primitives & domestic product lines on foreign primitives & domestic supply chain including hardware \\ \rowsep
A3 infrastructure operation & foreign-controlled & foreign infrastructure operated locally & national infrastructure on foreign hardware & national infrastructure with some domestic hardware & sovereign stack including domestic hardware \\ \rowsep
A4 governance and standards & pure taker & observer & active contributor & issues national profiles & authors a standard adopted by others \\ \rowsep
A5 research and talent & none & isolated researchers & university research groups & a national research program or institute & a sustained national ecosystem producing talent and IP \\
\bottomrule
\end{tabularx}
\end{table*}

\begin{table*}[t]
\centering\footnotesize
\caption{Indigenous post-quantum control codebook, crediting only the quantum-safe layer. The final
row is the external-dependency construct, scored in the opposite direction.}
\label{tab:cb_b}
\begin{tabularx}{\textwidth}{@{}l XXXXX@{}}
\toprule
Dim & 0 & 1 & 2 & 3 & 4 \\
\midrule
B1 indigenous PQC primitive & none & domestic research on foreign schemes & indigenous candidate or national call & indigenous scheme published or internationally submitted & nationally standardized indigenous algorithm \\ \rowsep
B2 PQC implementation & imports only & wraps a foreign library & domestic implementation of standard schemes & hardened or production implementations & full domestic toolchain \\ \rowsep
B3 PQC validation & foreign assurance & academic analysis & a national evaluation activity & national testing or certification & authoritative national program \\ \rowsep
B4 national PQC integration & follows a foreign schedule & adopts a foreign profile as-is & tailors a national profile & issues and enforces a national profile & full sovereign control \\ \rowsep
B5 PQC replaceability & waits for a foreign fix & swaps only vendor-shipped & documented agility on a foreign stack & agility with domestic implementation & agility with authorship or standardization \\ \rowsep
\addlinespace[1pt]
ED external dependency & self-sufficient national stack & largely domestic, minor foreign parts & foreign primitives offset by domestic implementation & foreign primitives and products, some domestic tooling & fully foreign-supplied stack \\
\bottomrule
\end{tabularx}
\end{table*}

\subsection{Aggregation and measurement}
Let $r_i$ ($i=1,\dots,5$) be the ordinal levels of a country's five readiness dimensions, $a_j$
($j=1,\dots,5$) the levels of its five capacity layers, $b_k$ ($k=1,\dots,5$) the levels of its five
post-quantum-control sub-variables, and $e$ the level of its single external-dependency cell, each
coded in $\{0,1,2,3,4\}$. Among the control sub-variables, $b_1$, $b_2$, and $b_3$ are the creation
layers of primitive design, implementation, and validation, and $b_4$ and $b_5$ are national
integration and replaceability. Readiness $R$ and indigenous cryptographic capacity ICC are the means
of their five levels, indigenous post-quantum creation Cr is the mean of the three creation levels, and
external dependency ED is the directly coded level,
\begin{align}
  R   &= \tfrac{1}{5}\textstyle\sum_{i=1}^{5} r_i, \label{eq:R}\\
  \mathrm{ICC} &= \tfrac{1}{5}\textstyle\sum_{j=1}^{5} a_j, \label{eq:icc}\\
  \mathrm{Cr}  &= \tfrac{1}{3}\,(b_1+b_2+b_3), \label{eq:cr}\\
  \mathrm{ED}  &= e, \label{eq:ed}
\end{align}
each therefore lying in $[0,4]$.
Cr is a descriptive score used for placement and comparison; the maker gate of \eqref{eq:gate} does not
operate on Cr, but on whether at least one of $b_1$, $b_2$, or $b_3$ reaches level three on a sustained
institutional basis. The post-quantum-control construct is coded across all five sub-variables B1 to B5, but the
two beyond creation, national integration $b_4$ and replaceability $b_5$, are reported per country
rather than folded into a single control index, because the gate and the maker type read the creation
layers and merging integration into the same axis would blur that distinction. Cr is therefore the
gate-relevant creation score and not a summary of the whole control construct, and the integration and
replaceability layers are read alongside it from the worksheet. The zero-to-four levels are ordinal anchors, and treating them as an equal-interval index
for these means is a modeling choice, the same one that composite policy indicators such as the human
development and national cyber indices make. Two features keep the choice from driving the results.
The primary maker gate is a threshold on a single creation dimension reaching level three, which needs
only the ordering of the levels and not their spacing, so it is invariant to any monotone rescaling.
And for the aggregated scores we test sensitivity directly, re-scoring under alternative monotone
spacings of the levels and reporting classification retention in Section~\ref{sec:results}. The model
deliberately does not merge capacity, control, and dependency into a single number, because that
merger is exactly the conflation the decomposition exists to undo, and a minimum across dimensions,
which would define a country by its weakest layer, was likewise set aside as discarding information
about countries strong on several layers and weak on one.

\subsection{A capacity-by-creation typology}\label{sec:typology}
Each country is placed on the plane of indigenous cryptographic capacity against indigenous
post-quantum creation, shown in Figure~\ref{fig:plane}, and the two are not collapsed. A post-quantum
maker clears the creation gate on a real capacity base, and its type, full-stack, research,
implementation, or validation, is read from which creation layers it holds. A capable, post-quantum-dependent country
holds high general capacity but has not built post-quantum control; this is where a country such as
Malaysia sits, high on capacity from a decade of national classical cryptography yet near zero on
post-quantum creation because its quantum-safe primitives are adopted, so describing it as capable but
post-quantum-dependent states what the evidence supports and corrects a reading that would call it a
maker on its classical base. A ready adopter has advanced its migration on readiness without either
capacity or post-quantum control, and a dependent country is low across the board; the classes are
listed in Table~\ref{tab:classes}. The rules reproduce
from the published scores. A country's class and, for a maker, its type are
\begin{equation}
  \mathrm{class}(c)=
  \begin{cases}
    \text{maker (typed)} & \text{if } \mathrm{maker}(c),\\
    \text{capable, PQC-dep.} & \text{else if } \mathrm{ICC}\ge 2,\\
    \text{ready adopter} & \text{else if } R\ge 2,\\
    \text{dependent} & \text{otherwise},
  \end{cases}
  \label{eq:class}
\end{equation}
\begin{equation}
  \mathrm{type}(c)=
  \begin{cases}
    \text{full-stack} & b_1\ge 3 \wedge b_2\ge 3,\\
    \text{research} & b_1\ge 3 \wedge b_2<3,\\
    \text{implementation} & b_1<3 \wedge b_2\ge 3,\\
    \text{validation} & b_1<3 \wedge b_2<3.
  \end{cases}
  \label{eq:type}
\end{equation}
Here $\wedge$ is logical conjunction, and the cases in \eqref{eq:class} are read from top to bottom, so
each ``else if'' applies only when every row above it does not; ``maker (typed)'' denotes a maker
labeled by its type from \eqref{eq:type}, and ``otherwise'' covers a country that clears neither the
capacity nor the readiness threshold.
The capacity and readiness cutoffs are the midpoint of the zero-to-four scale, so every class follows
from the maker gate of \eqref{eq:gate} together with two stated thresholds rather than a discretionary
judgment, and the gate verdict with the ICC and readiness scores reconstructs each country's class.
Readiness and
external dependency are reported alongside as separate axes, so the typology answers what a country
controls without conflating it with how far its migration has progressed or how much of the stack it
has borrowed.

The maker gate is a substantive threshold rather than a midpoint on a mean. It asks whether a country
has reached the level-three creation anchor on at least one of design, implementation, or validation,
and whether that capability rests on a sustained national institution rather than a single result, so
it separates a demonstrated national post-quantum capability from an isolated academic one. Countries
whose only qualifying evidence sits exactly at the anchor, or rests on a single low-confidence source,
are reported as boundary cases rather than definitive makers, and the uncertainty analysis of
Section~\ref{sec:results} quantifies how often each classification survives plausible re-coding.

\begin{table}[t]
\centering\small
\caption{The capacity-by-creation classes and their counts. Only makers clear the post-quantum
creation gate.}
\label{tab:classes}
\begin{tabularx}{\linewidth}{@{}Xc@{}}
\toprule
Class & Countries \\
\midrule
Full-stack maker (designs and implements) & 15 \\ \rowsep
Research maker (designs) & 5 \\ \rowsep
Implementation maker (implements) & 0 \\ \rowsep
Validation maker (validates) & 0 \\ \rowsep
Capable, post-quantum-dependent & 11 \\ \rowsep
Ready adopter & 1 \\ \rowsep
Dependent & 25 \\
\bottomrule
\end{tabularx}
\end{table}

\section{Methodology}\label{sec:method}
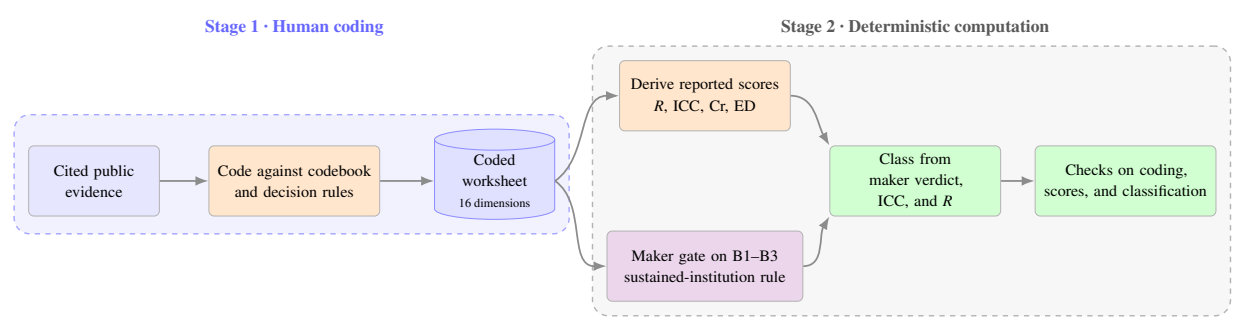
\begin{figure*}[t]
  \centering
  \resizebox{\textwidth}{!}{%
  \begin{tikzpicture}[font=\sffamily,
    >={Latex[length=2.6mm]},
    ar/.style={->,line width=1pt,black!45,rounded corners=2pt},
    ib/.style={rounded corners=3pt,draw=black!30,line width=0.5pt,align=center,
      text width=32mm,minimum height=14mm,inner sep=3pt,font=\small},
    orange/.style={ib,fill=orange!20},
    green/.style={ib,fill=green!18},
    purple/.style={ib,fill=violet!16},
    input/.style={ib,fill=blue!10,text width=24mm}]

    \node[input](ev) at (0,0) {Cited public\\evidence};
    \node[orange](code) at (4.0,0) {Code against codebook\\and decision rules};
    \node[cylinder,shape border rotate=90,aspect=0.28,draw=blue!45,fill=blue!12,
      minimum width=24mm,minimum height=18mm,align=center,font=\small,inner sep=1pt]
      (ws) at (8.0,0) {Coded\\worksheet\\{\scriptsize 16 dimensions}};

    \node[orange](agg) at (12.2,1.7) {Derive reported scores\\$R$, ICC, Cr, ED};
    \node[purple,text width=37mm](gate) at (12.2,-1.7) {Maker gate on B1--B3\\sustained-institution rule};
    \node[green](cls) at (16.4,0) {Class from\\maker verdict,\\ICC, and $R$};
    \node[green,text width=34mm](chk) at (20.6,0) {Checks on coding,\\scores, and classification};

    \begin{scope}[on background layer]
      \node[rounded corners=9pt,draw=blue!35,dashed,line width=0.8pt,fill=blue!5,
        fit=(ev)(code)(ws),inner sep=8pt]{};
      \node[rounded corners=9pt,draw=black!30,dashed,line width=0.8pt,fill=black!3,
        fit=(agg)(gate)(cls)(chk),inner sep=8pt]{};
    \end{scope}

    \draw[ar](ev.east)--(code.west);
    \draw[ar](code.east)--(ws.west);
    \draw[ar](ws.east) to[out=35,in=180] (agg.west);
    \draw[ar](ws.east) to[out=-35,in=180] (gate.west);
    \draw[ar](agg.east) to[out=0,in=120] (cls.north west);
    \draw[ar](gate.east) to[out=0,in=-120] (cls.south west);
    \draw[ar](cls.east)--(chk.west);

    \node[font=\bfseries,text=blue!60] at (4.0,3.05) {Stage 1 \textperiodcentered\ Human coding};
    \node[font=\bfseries,text=black!65] at (16.4,3.05) {Stage 2 \textperiodcentered\ Deterministic computation};
    \path (-2.1,-3.35) (22.7,3.55);
  \end{tikzpicture}}
  \caption{Two-stage RSCM scoring workflow, read left to right. All human judgment in the primary
  scoring pipeline ends at the coded worksheet; stage two is deterministic, so classes are computed
  rather than judged and the result is auditable rather than asserted. Blue band, human coding; gray
  band, deterministic computation.}\label{fig:method}
\end{figure*}

Because any capability index invites the objection that its scores are subjective, the
methodology is designed so that each value is evidence-anchored, auditable, and open to
challenge, and so that the results can be recomputed from the released coding. Figure~\ref{fig:method}
summarizes the workflow.

\subsection{The Scoring Pipeline}
The assessment is produced in two clearly separated stages, which confines every subjective judgment to
a transparent, cited record and makes the rest of the pipeline exactly reproducible. The separation is
deliberate and is what lets the paper be at once evidence-based and computationally reproducible.

The first stage is human coding. For each country a coder runs a structured search of public evidence
and, for every cell of the codebook, assigns an ordinal level from zero to four by matching the evidence
to that level's written anchor, recording alongside it an evidence-strength grade, a one-line
justification, and a citation to the source. All of the judgment in the primary scoring pipeline enters
here, at the level of one cell against a fixed rubric, where it is visible and contestable rather than
buried inside a composite. How far this stage reproduces across coders is exactly what the independent second coding of
Section~\ref{sec:results} measures.

The second stage is deterministic computation. A scripted pipeline takes the coded worksheet as its only
input and applies fixed rules with no further judgment. It aggregates the cell levels into the readiness,
capacity, creation, and dependency scores of \eqref{eq:R} to \eqref{eq:ed}, resolves the maker gate of
\eqref{eq:gate}, assigns each country a class and, for a maker, a type through \eqref{eq:class} and
\eqref{eq:type}, and runs the plausible-state bootstrap that yields the retention figures. The scoring
and classification rules are a pure function of the coded levels, so identical inputs always yield
identical scores and classes, and the plausible-state bootstrap is reproducible under a fixed random
seed, so every table, figure, and class in the paper can be regenerated from the released worksheet by
running the released scripts.

The consequence is a clean division of labor. Subjectivity is localized in the coding, which is cited
and audited and whose reproducibility is measured directly; the transformation from codes to results
carries none, so a reader who disputes a classification can trace it to the exact cells that produced it
and, on re-coding a cell, recompute the consequence deterministically. The companion explorer applies
the same second-stage rules in the browser, so an added or edited country is scored identically to the
paper.

\subsection{Country selection}
To keep the sample from appearing arbitrary, the selection is defined by an explicit sampling
frame and a set of inclusion and exclusion criteria specified in advance that are applied uniformly and
that another researcher can reconstruct.

The frame itself has two parts. The focus population is the set of countries of
Africa and the Arab world, taken as the membership of the African Union together with the League
of Arab States, which is the region the existing literature neglects and the study is designed to
characterize. To this is added a calibration set drawn from a distinct, objectively defined
population, namely the countries that are documented post-quantum actors, identified by the external
and checkable criterion of being recorded as having a national post-quantum program, mandate, or
standardization effort in public government-initiative trackers and the leading national-agency
publications \cite{gsma_pqc,uae_encryption,nsm10,bsi_pqc,anssi_pqc,kpqc,china_iccs,ncsc_uk,ccs_canada,asd_ism,csa_sg,india_nqm}.
The calibration set provides documented post-quantum actors outside the focus region, so the model can
be exercised across a wider range of capability configurations than a region-only study would reach.
The frame is defined by these two roles rather than by geography, so the assessment is
reported in Section~\ref{sec:results} across the six continents, and the Arab-world members of the
focus population fall within Africa for the countries of North Africa and within Asia for the
countries of the Gulf and the Levant.

Within each frame, a country is included if and only if it meets a
minimum-evidence threshold, namely that public evidence is sufficient to score a majority of the
coded dimensions from cited sources rather than from conservative defaults. The threshold is defined
on the availability of evidence rather than on the score a country would receive, so it is not
outcome-dependent, but it is not neutral either. Evidence availability is itself correlated with
capability and geography, since countries with stronger institutions, more published output, and more
English-indexed material are likelier to clear the threshold, and our search was principally in
English. The inclusion gate therefore selects toward the more capable and better-documented countries,
which narrows coverage of the least-resourced countries rather than tilts the scores of those
included, and we read the sample accordingly. The focus
population was swept in full against this gate, so every country of the African Union and the
League of Arab States with a documented national cryptographic footprint, meaning a national
public-key infrastructure or certification authority, a national cryptography or standards
agency with published output, documented cryptographic or quantum research, or recorded
participation in the international cryptography standards committee, is included. In the tables the
countries are grouped by class and, within a class, ordered by indigenous post-quantum creation for
readability rather than presented as a ranking of individual countries.

Conversely, a country in the frame that fails that same
threshold, typically a low-capacity or conflict-affected country with no public
cryptographic footprint, is excluded and assigned to a separate no-evidence category rather than
scored, because scoring it would mean fabricating a profile from defaults. Mauritania is an
instructive example, since it has a general national cybersecurity strategy but no public
cryptographic or post-quantum footprint, so nearly all of its dimensions could be scored only by
default; applying the threshold uniformly, it is placed in the no-evidence category rather than in
the sample, even though it lies within the focus population. This is the honest reason the regional
sample is a subset of its frame rather than a full census, and it is a transparent rule rather than
a discretionary choice; the excluded countries are not hidden but named as a category, and extending
coverage to them is identified as future work.

Both criteria are applied evenhandedly. Every included country, whether in the focus population or
the calibration set, is scored with the identical codebook, the identical source hierarchy, and
the identical decision rules, so no country receives favorable or unfavorable treatment by virtue of
its group. Because the frame and the threshold are stated explicitly, the sample can be reconstructed
and audited by anyone with access to the same public record. Reproducibility here has a precise and limited meaning.
Releasing the codebook, the coded worksheet, and the scripts makes the assessment computationally
reproducible, in that every table and figure can be regenerated from the coded data, and auditable, in
that every cell can be traced to its cited source. It does not establish that an independent coder
would derive the same levels from the raw evidence, which the moderate exact inter-coder agreement of
Section~\ref{sec:results} shows is not guaranteed. We claim the first two senses and are explicit that
the third, full reproducibility of the coding itself, is not demonstrated.

\subsection{Data sources and collection}
The assessment is built entirely from public evidence, gathered through a structured search
conducted for each country in turn and recorded in a per-country worksheet. For every country the search
covered the same classes of source, in a fixed order of preference. Primary sources were sought
first, namely national laws, regulations, decrees, and official strategy documents; the
publications, standards, and tools issued by national cyber-security and cryptography agencies;
the membership and contribution records of the international standards bodies; and the technical
documentation of national public-key infrastructures and certification authorities. Where primary
sources were silent, secondary sources were used and flagged as such, namely reputable technical
and industry press, vendor and telecommunications announcements, and the academic output of
national institutions and their researchers. Each country's national cyber agency, national
standards body, national public-key infrastructure operator, principal telecommunications
operators, and leading universities were checked in this way, and international bodies such as the
standards committees and regional consortia were consulted for cross-national signals.

Three properties of the collection process bear on the results. Primary evidence was preferred
over secondary throughout, and a score that rested on a single secondary source, or on a
structural inference such as the likely foreign origin of an undisclosed hardware security module,
was flagged in the worksheet rather than presented as established fact. The search was conducted
principally in English, with Arabic and French government sources consulted where they were
reachable, which is a limitation for exactly the regional core the paper focuses on, since some
national material in those languages is not indexed by open search; this limitation biases the
dependent countries toward lower scores and is revisited in Section~\ref{sec:threats}. Finally, the
evidence is a snapshot taken at a fixed date, because national mandates move quickly, and the
worksheet records the date and the source for every cell so that the assessment can be reproduced
and re-coded as the field advances. The complete worksheet records the level, its evidence-strength
grade, the justification, and the citation for every scored cell.

\subsection{Evidence and decision rules}
Every dimension is scored against the codebook from at least one cited public source, recorded in
the worksheet with a one-line justification, so that each cell can be traced to the artifact that
produced it. The sources are the public record a diligent analyst can reach, namely national
strategies and regulations, statements and tools published by national cyber and cryptography
agencies, the membership and contribution records of the international standards bodies, vendor
and telecommunications announcements, and the research output of national institutions, and where
a claim rests on a single source or on an inference it is flagged as such. Three rules are fixed
before scoring and applied uniformly. Where no public evidence supports a higher level, the lower
level is assigned and flagged as evidence-limited rather than inferred, which biases the
assessment toward understating capability. Announcements, plans, and pilots are capped below the
operational levels, so that a strategy document or a vendor pilot cannot be recorded as a
completed migration. And quantum key distribution is never scored as post-quantum deployment,
because it is a different response to the same threat. These rules do most of the work of keeping
the scoring disciplined, because the majority of contestable judgments in a fast-moving field
concern exactly the distinction between what has been announced and what is operational.

Two further distinctions are recorded because they bear on how the scores should be read. The first
separates the absence of evidence from evidence of absence, since no public document being found is
not the same as a public document establishing that a capability is missing. The worksheet therefore
tags each cell at one of three confidence bands, high for a level resting on primary evidence such as
a national law, standard, or agency publication, medium for one resting on reputable secondary
reporting, and low or evidence-limited for one resting on a single source, a structural inference
such as the presumed foreign origin of an undisclosed hardware security module, or a confirmed
absence in an incomplete record. The second distinction is that the model measures publicly
evidenced capability rather than complete national capability. Where cryptographic work is
classified, as it is for several countries with defense or signals-intelligence programs, the public
score understates the true capability, and this is stated for the affected countries rather than
concealed. The quantity the model estimates is therefore what a country can be shown to control from
the public record, which is the quantity a comparative and reproducible assessment can defend, and
the evidence-availability bias this introduces, which falls hardest on countries whose national material
is not indexed by open search, is revisited among the threats to validity.

\subsection{Reliability, robustness, and validity}
Reliability rests on the released coding worksheet, which records for every cell the level, its own
evidence-strength grade, a justification, and a citation, so that any reader can re-code a country
against the same anchors. An independent second coder re-coded a stratified sample of two hundred and
twenty cells across twenty countries, spanning the capacity, post-quantum-control, and
external-dependency dimensions and including all of the B1 to B5 control cells that decide the maker
gate. We report the agreement in full, the exact agreement, the agreement within one level, and the
quadratic-weighted Cohen's kappa, per construct rather than through a single figure, because
within-one agreement is a lenient target on a five-level scale, counting a level-two against a
level-three disagreement as agreement, and that is the disagreement the gate is sensitive to. Because
the gate is a threshold that turns a one-level cell disagreement into a whole reclassification, we also
report the agreement on the maker verdict itself, which is the quantity that bears on the headline
count. Robustness is examined in three ways, by re-scoring the aggregated capacity under alternative
monotone spacings of the ordinal levels, by testing the maker gate against a shift in its threshold,
and by a plausible-state bootstrap over the per-cell evidence grades that perturbs the most
judgment-dependent cells, since a single check can leave undisturbed a boundary that another would
move. All of these reliability and robustness results are reported in Section~\ref{sec:results}.

Validity is addressed by comparison against independent measures. The readiness axis is compared
with the Global Cybersecurity Index and the National Cyber Security Index \cite{itu_gci,ncsi}, and
the capacity and post-quantum-control constructs are checked against objective markers such as the
nationality of hardware security module manufacture, participation status in the relevant
international standards committee, and whether a country's institutions have designed a standardized
algorithm. Because some of these markers also inform the coding, the check is one of construct
alignment rather than fully independent validation, and it is reported as such rather than as proof.

Finally, the scores are treated as ordinal levels, and the classification foregrounds the maker gate
and the resulting classes rather than precise decimal rankings, because the ordinal construction does
not support arithmetic on small differences and because the policy conclusions depend on class
membership rather than on the exact position within a class. The maker gate is a threshold on the
ordering of a single creation level, so it depends only on that ordering and is invariant to any
monotone rescaling of the levels. The codebook, the evidence worksheet, and the scoring scripts
are released as supplementary material so that every table and figure can be reproduced and any
country re-scored against the same anchors.

\subsection{Coding examples}
Four coded examples show how the codebook turns evidence into a score. Each gives the dimension
and country, the public evidence, the codebook level that evidence matches, and the resulting score.
\begin{enumerate}
  \item \textbf{Strategy and mandate, United Arab Emirates.} The evidence is a binding national
  encryption policy that requires transition plans and cryptographic discovery
  \cite{uae_encryption}. It matches the level-four indicator, a binding legal instrument with an
  enforced deadline, so the score is four.
  \item \textbf{Standards adoption, Saudi Arabia.} The evidence is a national cryptographic standard
  that is operational but classical and treats PQC only as an appendix \cite{nca_ncs}. It matches
  the pilot level rather than the mandated-profile level, so the score is two.
  \item \textbf{Infrastructure and dependency, Kenya.} The evidence is a national public-key
  infrastructure whose platform and hardware are supplied by a foreign vendor \cite{kenya_npki}. It
  matches the national-infrastructure capacity level, a score of two, while the foreign supply of
  the platform raises the external-dependency construct.
  \item \textbf{Indigenous post-quantum primitive, Senegal.} The evidence is a code-based scheme its
  researchers authored and submitted to an international competition \cite{dags,dags_sn}. It matches
  the indigenous-primitive level, a creation score of three, but as a single university result it
  does not meet the sustained-institution requirement of the maker gate.
\end{enumerate}
The examples also show the conservative rule at work, since where a higher level would require
evidence that no public source provides, the lower level is assigned and flagged rather than
inferred.

\section{Results}\label{sec:results}
The decomposition separates what a single sovereignty score conflates. Table~\ref{tab:scores}
reports, for all fifty-seven countries, readiness $R$, indigenous cryptographic capacity ICC, the
indigenous post-quantum creation score, external dependency ED, and the class, and
Figure~\ref{fig:plane} shows the plane of capacity against creation. The first result is that capacity
and post-quantum creation are distinct axes: several countries with high general capacity sit near
zero on post-quantum creation, which is the configuration a one-dimensional score hides and by which
strong classical capability could be read as post-quantum control.

\begin{table*}[t]
\centering\small
\setlength{\tabcolsep}{4pt}
\caption{Readiness $R$, capacity ICC, creation Cr, dependency ED, and class for all fifty-seven
countries, grouped by class and, within a class, ordered by creation Cr for readability; the order is
not a ranking of countries. Capable abbreviates capable but post-quantum-dependent; a dagger marks a
boundary case; full per-cell scores and citations are in the released worksheet.}
\label{tab:scores}
\begin{tabular}{@{}>{\raggedright\arraybackslash}p{1.95cm} cccc >{\raggedright\arraybackslash}p{1.7cm} c >{\raggedright\arraybackslash}p{1.95cm} cccc >{\raggedright\arraybackslash}p{1.7cm}@{}}
\toprule
Country & $R$ & ICC & Cr & ED & Class & & Country & $R$ & ICC & Cr & ED & Class \\
\midrule
United States & 3.8 & 4.0 & 3.7 & 1 & Full-stack & & Malaysia & 2.2 & 2.8 & 0.3 & 4 & Capable \\ \rowsep
Germany & 3.0 & 3.6 & 3.3 & 2 & Full-stack & & Ethiopia & 0.8 & 2.4 & 0.0 & 4 & Capable \\ \rowsep
Ukraine & 2.0 & 3.4 & 3.3 & 1 & Full-stack & & Singapore & 2.6 & 1.6 & 1.0 & 4 & Ready adopter \\ \rowsep
Russia & 2.0 & 3.8 & 3.0 & 1 & Full-stack & & Senegal & 0.4 & 1.6 & 2.0 & 3 & Dependent \\ \rowsep
France & 3.0 & 3.6 & 3.0 & 2 & Full-stack & & Iraq & 0.6 & 1.8 & 0.7 & 4 & Dependent \\ \rowsep
South Korea & 2.6 & 3.2 & 3.0 & 2 & Full-stack & & Tunisia & 1.0 & 1.6 & 0.7 & 4 & Dependent \\ \rowsep
Netherlands & 2.6 & 3.0 & 3.0 & 1 & Full-stack & & Jordan & 1.2 & 1.6 & 0.7 & 4 & Dependent \\ \rowsep
United Kingdom & 3.0 & 2.4 & 3.0 & 2 & Full-stack & & Lebanon & 0.8 & 1.6 & 0.7 & 4 & Dependent \\ \rowsep
China & 3.0 & 3.8 & 2.7 & 1 & Full-stack & & South Africa & 1.2 & 1.6 & 0.7 & 4 & Dependent \\ \rowsep
Switzerland & 2.4 & 3.4 & 2.7 & 2 & Full-stack & & New Zealand & 1.4 & 1.8 & 0.3 & 4 & Dependent \\ \rowsep
Taiwan & 2.0 & 2.8 & 2.7 & 2 & Full-stack & & Indonesia & 1.6 & 1.8 & 0.3 & 4 & Dependent \\ \rowsep
Canada & 2.6 & 2.8 & 2.7 & 2 & Full-stack & & Egypt & 1.0 & 1.8 & 0.3 & 4 & Dependent \\ \rowsep
UAE & 3.2 & 2.6 & 2.7 & 2 & Full-stack & & Morocco & 1.0 & 1.8 & 0.3 & 4 & Dependent \\ \rowsep
Belgium & 1.8 & 3.4 & 2.3 & 2 & Full-stack & & Cameroon & 0.2 & 0.8 & 0.3 & 4 & Dependent \\ \rowsep
Austria & 1.0 & 2.8 & 2.3 & 2 & Full-stack & & Algeria & 0.4 & 1.4 & 0.0 & 4 & Dependent \\ \rowsep
Japan & 2.0 & 3.2 & 3.0 & 2 & Research & & Kenya & 0.4 & 1.4 & 0.0 & 4 & Dependent \\ \rowsep
Turkey$^{\dagger}$ & 1.4 & 3.6 & 2.3 & 3 & Research & & Bahrain & 1.0 & 1.2 & 0.0 & 4 & Dependent \\ \rowsep
Vietnam$^{\dagger}$ & 1.4 & 2.6 & 2.3 & 2 & Research & & Nigeria & 0.4 & 1.2 & 0.0 & 4 & Dependent \\ \rowsep
Brazil$^{\dagger}$ & 1.8 & 2.4 & 2.3 & 2 & Research & & Rwanda & 0.4 & 1.2 & 0.0 & 4 & Dependent \\ \rowsep
Australia$^{\dagger}$ & 2.4 & 2.0 & 2.3 & 3 & Research & & Mauritius & 1.0 & 1.2 & 0.0 & 4 & Dependent \\ \rowsep
Estonia & 2.4 & 2.8 & 1.7 & 3 & Capable & & Uganda & 0.6 & 1.2 & 0.0 & 4 & Dependent \\ \rowsep
Argentina & 0.8 & 2.2 & 1.7 & 2 & Capable & & Cote d'Ivoire & 0.2 & 1.2 & 0.0 & 4 & Dependent \\ \rowsep
Israel & 2.4 & 3.4 & 1.3 & 3 & Capable & & Oman & 0.6 & 1.0 & 0.0 & 4 & Dependent \\ \rowsep
India & 2.4 & 2.2 & 1.3 & 3 & Capable & & Kuwait & 0.6 & 0.8 & 0.0 & 4 & Dependent \\ \rowsep
Mexico & 0.6 & 2.0 & 1.3 & 3 & Capable & & Ghana & 0.4 & 0.8 & 0.0 & 4 & Dependent \\ \rowsep
Iran & 0.8 & 2.4 & 1.0 & 4 & Capable & & Namibia & 0.4 & 0.8 & 0.0 & 4 & Dependent \\ \rowsep
Qatar & 1.8 & 2.0 & 1.0 & 4 & Capable & & Tanzania & 0.6 & 0.8 & 0.0 & 4 & Dependent \\ \rowsep
Pakistan & 0.8 & 2.0 & 1.0 & 4 & Capable & & Botswana & 0.4 & 0.8 & 0.0 & 4 & Dependent \\ \rowsep
Saudi Arabia & 1.8 & 2.4 & 0.7 & 4 & Capable & &  & & & & & \\ \rowsep
\bottomrule
\end{tabular}
\end{table*}

\subsection{The post-quantum makers}
Twenty countries clear the post-quantum creation gate on a sustained institutional basis, and the
typology does not equate them. Fifteen are full-stack makers that both design and implement
post-quantum cryptography, among them the United States, China, Germany, France, South Korea, the
United Kingdom, the Netherlands, Switzerland, Russia, Ukraine, Taiwan, Canada, Belgium, Austria, and
the United Arab Emirates. The type is a qualifying threshold rather than a ranking of depth, and the
full-stack makers span a wide range on both axes. The United States sits at the top with a creation
score of 3.7 and a broad national ecosystem, while the United Arab Emirates enters the same type at 2.7
on a much narrower base, its design and implementation layers clearing the gate without matching that
scale. The shared label certifies that design and implementation both reach the anchor, not equal
breadth, which the creation score and the capacity axis carry separately. Five are research makers that
design or analyze post-quantum cryptography
without a domestic production base, Japan together with the boundary cases Australia, Vietnam,
Brazil, and Turkey. The remaining two types are empty here. No country is a pure implementation maker,
which implements without originating a primitive, and none is a validation maker, which clears the gate
through validation alone; both types are defined for completeness because each is a distinct route to
control, and their absence, every maker in the sample holding at least a design or a design-and-build
profile, is itself informative. Table~\ref{tab:makers} gives
the type, the post-quantum creation score with its uncertainty interval, and the retention of maker
status under re-coding.

The corrected attribution matters here. The United States is credited for standardization authority
and full-stack engineering rather than for authoring the primitives, which international academic teams
designed. The United Arab Emirates enters as a full-stack maker on a genuine post-quantum basis, on
demonstrated design and implementation \cite{tii_pqclib} rather than on its binding national encryption
policy, which is credited separately under national integration. Boundary status has an explicit rule that a
reader can reconstruct from the worksheet: a maker is a boundary case when it clears the gate on a
single creation cell at exactly the level-three anchor, with no second qualifying layer above it.
Four makers meet this rule, Turkey, Vietnam, Brazil, and Australia, each carried by an indigenous
primitive design and nothing else at level three. This structural fragility is distinct from
evidential fragility, which the retention column reports separately, and the two do not coincide.
Australia clears on a single cell but that cell rests on high-grade primary evidence, so its
bootstrap retention is complete, whereas Turkey, Brazil, and Vietnam clear on a single cell graded
medium, so their retention is lower. A boundary flag therefore marks a country that a re-coding of one
cell could move, and the retention figure says how likely the evidence makes that move. The gate also
excludes a case
that a looser reading would admit. Argentina operates a genuine domestic secure-communications
system, but its only claimed post-quantum content is a proprietary, undisclosed scheme that has never
been published or independently analyzed, so no publicly verifiable post-quantum-specific capability
reaches the gate, and Argentina is classed capable but post-quantum-dependent rather than a maker.
The exclusion is deliberate, since crediting an undisclosed proprietary claim would reintroduce
exactly the over-attribution the gate is built to prevent.

\begin{table}[t]
\centering\small
\caption{The twenty post-quantum makers with type, creation score, its ninety-percent interval, and
the share of bootstrap draws that retain maker status. A dagger marks a boundary maker.}
\label{tab:makers}
\setlength{\tabcolsep}{4pt}
\begin{tabular}{@{}>{\raggedright\arraybackslash}p{2.0cm} >{\raggedright\arraybackslash}p{1.95cm} c c r@{}}
\toprule
Country & Type & Cr & Interval & Retention \\
\midrule
United States & Full-stack & 3.7 & 3.67--3.67 & 100\% \\ \rowsep
Germany & Full-stack & 3.3 & 3.0--3.67 & 100\% \\ \rowsep
Ukraine & Full-stack & 3.3 & 3.0--3.67 & 100\% \\ \rowsep
Russia & Full-stack & 3.0 & 2.67--3.33 & 100\% \\ \rowsep
France & Full-stack & 3.0 & 2.67--3.33 & 100\% \\ \rowsep
South Korea & Full-stack & 3.0 & 2.67--3.33 & 100\% \\ \rowsep
Netherlands & Full-stack & 3.0 & 2.67--3.33 & 100\% \\ \rowsep
United Kingdom & Full-stack & 3.0 & 2.67--3.33 & 100\% \\ \rowsep
China & Full-stack & 2.7 & 2.33--3.0 & 100\% \\ \rowsep
Switzerland & Full-stack & 2.7 & 2.33--3.0 & 100\% \\ \rowsep
Taiwan & Full-stack & 2.7 & 2.33--3.0 & 100\% \\ \rowsep
Canada & Full-stack & 2.7 & 2.33--3.0 & 100\% \\ \rowsep
UAE & Full-stack & 2.7 & 2.33--3.0 & 100\% \\ \rowsep
Belgium & Full-stack & 2.3 & 2.0--2.67 & 100\% \\ \rowsep
Austria & Full-stack & 2.3 & 2.0--2.67 & 100\% \\ \rowsep
Japan & Research & 3.0 & 2.67--3.33 & 100\% \\ \rowsep
Turkey & Research & 2.3 & 1.67--3.0 & 92\% \\ \rowsep
Vietnam & Research & 2.3 & 1.67--3.0 & 87\% \\ \rowsep
Brazil & Research & 2.3 & 1.67--3.0 & 87\% \\ \rowsep
Australia & Research & 2.3 & 1.67--3.0 & 100\% \\ \rowsep
\bottomrule
\end{tabular}
\end{table}

\subsection{Capacity without post-quantum control}
The correction is sharpest for the countries that hold high general cryptographic capacity but have
not built post-quantum control. Eleven countries are capable but post-quantum-dependent: Malaysia,
Saudi Arabia, Qatar, Israel, India, Iran, Estonia, Argentina, Mexico, Pakistan, and Ethiopia.
Malaysia is the clearest
case, with an indigenous cryptographic capacity above the midpoint from a decade of national classical
cryptography but adopted post-quantum primitives and near-zero post-quantum creation. The same logic
applies to the Gulf countries and to Israel and India, whose strong general and academic bases have not
yet produced controlled national post-quantum capability.

\subsection{The dependent majority}
\begin{figure*}[t]
  \centering
  \includegraphics[width=0.94\textwidth]{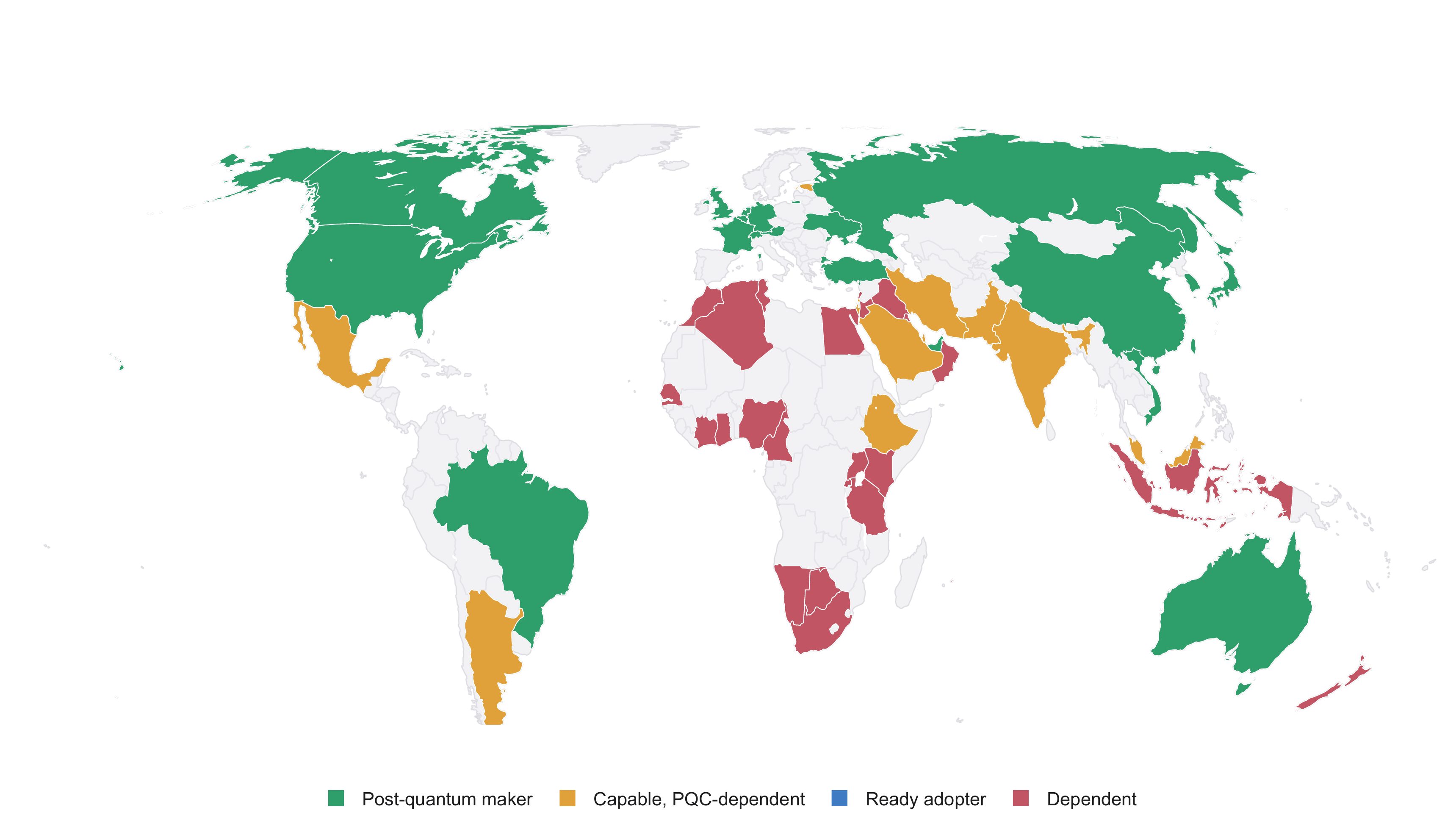}
  \caption{Distribution of the classes across the assessed countries; makers concentrate in Europe,
  North America, and East Asia. Drawn on the equal-area Equal Earth projection \cite{un_equalearth};
  the sample is purposive, not a world census.}
  \label{fig:world}
\end{figure*}

Twenty-five countries are dependent, low on capacity and on post-quantum creation alike, and one,
Singapore, is a ready adopter that has advanced its migration without either capacity or control.
Figure~\ref{fig:world} maps the classes across the assessed countries.

\subsection{Regional distribution}
Reporting the assessment across the six continents shows a steep capability gradient that the
focus-and-comparator split of the sampling frame conceals. Table~\ref{tab:regions} gives the mean of
each construct and the class mix by continent. Europe and North America lead, with the highest
indigenous capacity and post-quantum creation and the lowest external dependency, and between them
they hold twelve of the twenty makers. Asia is the widest-spread continent and the pivot of the
picture, since its twenty-one countries include six makers in East Asia and the Gulf alongside seven
capable but post-quantum-dependent countries and seven dependent ones, so a single continental mean
averages over China, Japan, South Korea, and Taiwan at one end and the low-capacity countries of
South and West Asia at the other. South America and Oceania are each represented by only
two countries, so their per-continent means are reported for completeness and carry no interpretive
weight; each pairs one maker at the two-to-three creation boundary with one lower-capability country,
and a two-country average is too thin to place on the gradient. Africa is the floor of the distribution, with the lowest capacity
and creation and the highest external dependency of any continent, seventeen of its eighteen countries
dependent and none a maker, the one exception being a single capable but post-quantum-dependent
country.

Two features of the sample bound this reading. The assessed set is a purposive sample of documented
cryptographic actors rather than a census, so each continent is represented by its better-documented
members. Because evidence availability is associated with institutional capacity, the assessed sample
may overrepresent better-documented and more capable countries, and the direction and magnitude of the
resulting bias cannot be established for countries that were not scored. The gradient is therefore a
statement about the assessed countries rather than an unbiased description of the world.

\begin{table}[t]
\centering\small
\setlength{\tabcolsep}{4pt}
\caption{Per-continent means of $R$, ICC, Cr, and ED with the class mix (makers / capable but
post-quantum-dependent / ready adopters / dependent), for the fifty-seven countries. Russia and Turkey
count in Europe, Mexico in North America.}
\label{tab:regions}
\begin{tabular}{@{}l r cccc c@{}}
\toprule
Continent & $n$ & $R$ & ICC & Cr & ED & Classes \\
\midrule
Europe & 11 & 2.2 & 3.3 & 2.7 & 1.9 & 10/1/0/0 \\ \rowsep
North America & 3 & 2.3 & 2.9 & 2.6 & 2.0 & 2/1/0/0 \\ \rowsep
South America & 2 & 1.3 & 2.3 & 2.0 & 2.0 & 1/1/0/0 \\ \rowsep
Asia & 21 & 1.7 & 2.2 & 1.3 & 3.3 & 6/7/1/7 \\ \rowsep
Oceania & 2 & 1.9 & 1.9 & 1.3 & 3.5 & 1/0/0/1 \\ \rowsep
Africa & 18 & 0.6 & 1.3 & 0.2 & 3.9 & 0/1/0/17 \\ \rowsep
\bottomrule
\end{tabular}
\end{table}

\subsection{Uncertainty and robustness}
The independent second coding measures how reliably the coding reproduces, and the
post-quantum-control cells that decide the gate are its most reliable part. Across the two hundred and
twenty re-coded cells, exact agreement was fifty-three percent, agreement within one level ninety
percent, and the quadratic-weighted kappa 0.65; the five control cells reached 0.74 and the three gate
cells B1 to B3 reached 0.71, with sixty-three percent exact and ninety-three percent within-one
agreement, above the capacity cells at 0.39 and the external-dependency cells at 0.58, with
Table~\ref{tab:irr} reporting the full per-construct agreement. Because the gate is a threshold, the decisive quantity is
the agreement on the maker verdict itself, and here the two coders agreed for fourteen of the twenty
countries and disagreed on six, every disagreement falling among countries whose creation cells sit at
the two-to-three threshold rather than in the core. Two of the four formal boundary makers, Vietnam
and Brazil, fell below the gate for the second coder while Turkey and Australia held, and four
non-makers just under the gate, Malaysia, Israel, Argentina, and Singapore, crossed above it, while the
nine makers that clear the gate on more than one cell or on high-grade evidence never moved. The maker
classification is therefore reliable in its core and uncertain along the two-to-three threshold in both
directions, so the twenty-maker count should be read as a robust core of roughly a dozen makers with a
contested margin that the formal boundary flag marks only on the maker side.

\begin{table}[t]
\centering\small
\setlength{\tabcolsep}{4pt}
\caption{Inter-coder reliability from the second coding, by construct. Exact is exact agreement, W1
agreement within one level, and $\kappa_w$ the quadratic-weighted Cohen's kappa.}
\label{tab:irr}
\begin{tabular}{@{}l r r r r@{}}
\toprule
Construct & $n$ & Exact & W1 & $\kappa_w$ \\
\midrule
All coded cells & 220 & 53\% & 90\% & 0.65 \\ \rowsep
Cryptographic capacity (A1--A5) & 100 & 46\% & 84\% & 0.39 \\ \rowsep
Post-quantum control (B1--B5) & 100 & 60\% & 94\% & 0.74 \\ \rowsep
Maker-gate cells (B1--B3) & 60 & 63\% & 93\% & 0.71 \\ \rowsep
External dependency (ED) & 20 & 50\% & 95\% & 0.58 \\
\bottomrule
\end{tabular}
\end{table}

Each cell carries an evidence-strength grade, and its plausible-state set is bounded at one level,
with per-grade perturbation probabilities that reproduce the ninety percent overall within-one-level
agreement between the two coders. We are cautious about what this calibration can and cannot claim.
Within-one agreement is a lenient target, since it counts a level-two against level-three disagreement
as agreement, and that is exactly the disagreement the maker gate is sensitive to, so a bootstrap tuned
to within-one understates the uncertainty at the gate. The bootstrap is therefore a lower bound on
gate uncertainty, and we read its retention figures alongside the direct gate-sensitivity evidence,
the boundary makers and the countries that sit just below the gate, rather than as a sufficient
account on their own. Over a four-thousand-draw bootstrap across these countries, maker status is
retained in ninety-six percent of draws on average, and eighteen of the twenty makers retain the label
in at least ninety percent of draws. The full-stack core, the United States, China, Germany, France, South Korea, and the
United Kingdom, retains it in every draw. The countries whose status is most sensitive are those
sitting just below the creation gate, Estonia, Israel, Argentina, Mexico, and India, which cross into
the makers only when a low-confidence creation cell is drawn upward, and the four boundary makers,
whose intervals Table~\ref{tab:makers} reports. Because the maker gate is a threshold on the ordering
of a single creation level rather than on the spacing of the scale, it is invariant to any monotone
rescaling of the levels, and re-scoring the aggregated capacity under alternative monotone spacings
leaves the class of every country unchanged.

The bootstrap and the second coding agree that the maker core is stable and locate all the movement at
the gate threshold. The bootstrap retention nonetheless understates the margin, because inter-coder
disagreement at the two-to-three line, which the second coding measures directly, is larger than the
evidence-grade perturbation the bootstrap applies.

\section{Discussion}\label{sec:discussion}
The assessment supports a single structural reading once sovereignty is decomposed. Control of the
post-quantum layer is concentrated in the makers that design, implement, or validate quantum-safe
cryptography on a sustained national basis, and it is a narrower set than the countries with strong
general cryptographic capacity. The decomposition is the paper's central correction: a one-dimensional
sovereignty score merges general capacity, post-quantum control, and external dependency, and it is
that merger, not any single weight, that allowed classical capability and governance structure to be
read as post-quantum sovereignty. Separating the constructs shows that many capable countries have not
yet built post-quantum control, that using a foreign standardized primitive is not by itself a loss of
sovereignty when a country controls implementation and replacement, and that authorship is one strong
route to control among several.

\subsection{Case studies}
Five cases make the typology concrete, one for each way a country meets or misses the gate.

The United States is a full-stack maker that designs, implements, validates, and
controls national post-quantum integration, and it standardized the primary algorithms through an open
international competition. It is credited for standardization authority and full-stack engineering
rather than for authoring the primitives, which international academic teams designed, and the
correction sharpens rather than diminishes its position, since it holds every creation and control
layer.

The United Arab Emirates is a full-stack maker on a genuine post-quantum basis. It
enters through its national research institute, which co-designed a standardized mechanism and
produced a post-quantum library with a corresponding hardware implementation integrated into several
secure-communication products \cite{tii_pqclib}, so its design and implementation layers both reach
the level-three anchor. The binding national encryption policy is credited separately under national
integration rather than as the basis for the maker classification, which rests on demonstrated
post-quantum creation, not on the governance and infrastructure that a general sovereignty score would
have credited, which is the distinction the decomposition is built to draw.

Malaysia, by contrast, is capable but post-quantum-dependent. It holds high general cryptographic
capacity from a decade of national classical cryptography, a national trusted-algorithm project, and a
domestic research community, yet its post-quantum primitives are adopted and its post-quantum creation
is near zero. It is capable but post-quantum-dependent, not a maker, and treating its classical
capability as evidence of post-quantum sovereignty is exactly the conflation the decomposition
prevents.

Belgium and Austria are full-stack makers on a research-led base. Two of the strongest
cryptographic-design nations, they both design and implement post-quantum cryptography, Belgium through the
authorship of SABER and masked, side-channel-resistant implementations at COSIC, Austria through the
co-authorship of the standardized SPHINCS+ and the KaLi lattice coprocessor at TU Graz. They clear the
gate on both the design and the implementation layer and are full-stack makers, though their
implementation base is research-led rather than a broad national industry, which is the difference
between clearing the gate on both creation layers and doing so at the scale of the largest makers.

Senegal, finally, is a dependent country with an isolated result. A single university group's
code-based submission is a real academic contribution but not a sustained national capability, so
Senegal does not clear the gate and is classed dependent, an illustration of the sustained-institution
requirement that keeps an isolated result from being read as national sovereignty.

\subsection{Broader implications}
The pattern connects the quantum transition to the wider debate on digital sovereignty and
supply-chain security. A country that imports its algorithms, its hardware security modules, and its
migration tooling places the foundation of its digital order in a supply chain it does not control, and
the transition is an occasion to reconsider that dependency in the cryptographic supply chain. The
concentration of post-quantum creation in a few makers also raises the prospect of a multi-standard
world, in which a country must choose between algorithm families backed by different powers, and there
the ability to design, or at least to independently implement and evaluate, a scheme becomes a strategic
capability rather than a technical detail. Decomposing sovereignty is intended to make this dimension
legible to the ministries that will make procurement and policy decisions over the next decade.

The framing has limits that should be stated. It compresses a continuous and multi-layered reality
into a small set of classes, and although the creation profile and the external-dependency axis
recover much of the within-class magnitude, the class itself remains a discrete kind rather than a
point on a continuum. It also treats the layers of control as separable when in practice they
interact, and it says nothing about the quality or trustworthiness of a given maker's cryptography,
only about who controls it. The model measures control and dependency, not cryptographic security, so
domestic capability is not treated as more secure and foreign capability as less secure; a
well-scrutinized foreign open standard may be more trustworthy than an unproven domestic one, and a
country may rationally adopt it for interoperability, cost, assurance, or efficiency. The vocabulary is
operational rather than normative, and none of the class labels implies political inferiority. What the
model claims is narrow, that ignoring the distinction between general capacity, post-quantum control,
and dependency leads a country to mistake a foreign-supplied migration for a sovereign one.

\subsection{Scenario and historical parallel}
The practical meaning of borrowed post-quantum sovereignty is clearest under a concrete scenario.
Suppose that within the coming decade a cryptanalytic advance weakens the structured-lattice family on
which the primary standardized algorithms rest. A full-stack maker would respond from a position of
control, drawing on its own analysts to assess the advance, its design capacity to propose a revision,
and its implementations to reissue software and hardware, and it would have a seat where the corrected
standard is decided. A capable but post-quantum-dependent country, however strong its general
cryptography, would depend on a maker to detect the problem, design the fix, and schedule the
correction, and its migration would proceed on the terms that others set. The dependent countries
would be furthest back in the queue, reliant on foreign vendors to ship a remediation they could
neither evaluate nor accelerate. The scenario is hypothetical, but it is the exact contingency for
which post-quantum control is insurance, and it shows why measuring that control is not academic.

The pattern has historical parallels in other foundational technologies. Countries that adopted
foreign telecommunications, satellite, or industrial-control standards without domestic capacity found
that their apparent modernization carried a durable dependence that surfaced whenever the supplying
relationship changed. Cryptography differs in degree rather than kind, but the degree matters, because
cryptography sits beneath everything else and because the confidentiality lifetimes at stake are
unusually long. In this respect the quantum transition follows a familiar pattern, in which countries
that use a foundational shift to build domestic capacity retain more autonomy afterward than countries
that meet it mainly through procurement, and the resulting difference in dependence persists well beyond
the transition.

\section{Limitations and Threats to Validity}\label{sec:threats}
We state the principal limitations first, then take the threats to validity by type.

\subsection{Limitations}
Several limitations bound what the assessment can claim, and each is analyzed in detail where it
arises. The first and most consequential is the reliability of the maker classification at its
boundary. The gate is a threshold, and although the creation cells that feed it are coded reliably, an
independent second coder moved the maker verdict for six of twenty re-coded countries, every one of
them sitting at the two-to-three gate threshold rather than in the core, so the twenty-maker count is
best read as a robust core of roughly a dozen makers together with a contested margin rather than an
exact figure, as Sections~\ref{sec:method} and~\ref{sec:results} report. The second is that the central construct, indigenous post-quantum control, has no independent
external criterion, so the evidence for it is internal robustness and discriminant separation from
readiness rather than construct validity in the strict sense. The third is that the capacity axis is the least reliable component of the
coding and its scores should be read as indicative. The fourth is that the sample is a purposive set of
documented actors rather than a census, and because evidence availability is correlated with capability
and with English-language indexing, coverage is thinner for the least-resourced countries. The fifth,
consequential for a framework that presents itself as a measurement instrument, is that its dimensions
and ordinal anchors are conceptually grounded and reliably applied but not established as content-valid
through a formal derivation, so the model is best read as a proposed and operationalized instrument
rather than a validated one, as the construct-validity discussion below sets out. Two narrower
limits complete the list, that the assessment scores publicly evidenced capability and so understates
classified national programs, and that it is a snapshot of a fast-moving field. The remaining
subsections take the corresponding threats to validity in turn.

\subsection{Construct validity}
The most consequential construct decision is the decomposition itself, the claim that general
cryptographic capacity, post-quantum control, and external dependency are distinct constructs
rather than one sovereignty axis. Conflating them would let classical capability and governance
authority stand in for post-quantum control, and the maker gate exists to hold the post-quantum
construct to demonstrated post-quantum capability. We mitigate the residual risk by grounding each dimension in
an observable national artifact rather than a latent trait, so that a reader who disputes a
dimension can inspect exactly what it measures, and by reporting the capacity, creation, and
dependency scores separately so that no single number hides a country's true position. The framing
still compresses a multi-layered reality into a small set of classes, which is a deliberate
simplification whose cost is discussed in Section~\ref{sec:discussion}.

A distinct construct-validity question concerns not whether the three axes separate but whether each is
measured by the right dimensions and levels, and here the framework is operationalized and
reliability-tested rather than shown to be content-valid. The readiness dimensions are adapted from the
national post-quantum migration and maturity literature, and the control dimensions follow the working
definition of sovereignty as the capacity to design, implement, validate, integrate, and replace the
cryptography a nation deploys, but neither set was established through a formal content-validity
procedure such as expert elicitation, a Delphi study, or an empirical dimension reduction. We therefore
do not claim that the five readiness dimensions, the five capacity layers, or the five control
sub-variables are exhaustive or non-overlapping, and a defensible alternative might promote procurement,
budgeting, or crypto-agility to a dimension of its own. The zero-to-four scale follows the common
maturity-model convention of five ordered levels, and although its anchors are stated so a reader can
judge each step, the number of levels and the placement of the boundaries are design choices rather
than empirically derived thresholds, as is the equal-interval treatment of the levels in the reported
means. The same holds for the two classification cutoffs beyond the gate, the scale midpoint for
capacity and readiness, which are stated conventions; only the maker gate at level three carries a
substantive rationale, the transition from activity to demonstrated national creation. Establishing the
content validity of the dimensions and anchors, through expert elicitation or empirical dimension
analysis on a larger sample, is the most important measurement-development step left open, and until it
is done the framework should be read as a proposed, operationalized, and reliability-tested measurement
model rather than a validated instrument.

\subsection{Measurement validity}
The assessment rests on public evidence, so an absent verdict indicates that no public source
was found rather than that a capability does not exist, and government material in Arabic and
French may hold detail that open search does not index. This risk is largest for the dependent
countries, and it cuts against them, because a capability that exists but is not publicly documented
is scored low and can make a country look more dependent than it is. We do not treat that missing
evidence as support for the dependent finding, since absence of evidence is not evidence of
dependence. What supports the finding is positive evidence rather than absent evidence: the dependent
classifications rest on documented foreign provenance, such as foreign-manufactured hardware security
modules, adopted foreign primitives, and nationally operated public-key infrastructure supplied by
foreign vendors, so the reading is anchored in what the record shows a country uses rather than in
what it fails to show a country builds. The residual risk is that a country with undocumented
indigenous capability is placed too low, which we flag for the affected countries rather than resolve.
The most judgment-dependent cells are deployment maturity and
the post-quantum creation cells that decide the maker gate, and these carry the lowest evidence
grades. The plausible-state bootstrap shows that class assignments survive perturbation of these
cells, and the sustained-institution requirement of the gate constrains them, but they would firm
up with per-sector production data that is not publicly available for most countries.

\subsection{External validity, sampling, and time}
Beyond the construct-alignment check of Section~\ref{sec:results}, the readiness axis shows convergent
validity against two independent third-party indices that are used nowhere in the coding. Over the forty-nine
assessed countries that the National Cyber Security Index \cite{ncsi} covers, Spearman's rank
correlation between readiness and the index is 0.70 at a significance below the one percent level;
against the ITU Global Cybersecurity Index \cite{itu_gci}, a coarser five-tier commitment measure
that also covers most of the countries the first index omits, readiness again correlates, at 0.48 and
again below the one percent level. The readiness axis therefore tracks two independently constructed
measures of national cyber posture. The post-quantum construct behaves differently and in the
expected way. Indigenous post-quantum creation correlates with the fuller first index at 0.63, since
capable countries tend to score well on both, but it is essentially uncorrelated with the commitment
tiers of the second index, at 0.22 and not significant, so post-quantum control is not a restatement
of a country's cyber-security commitment. Its distinctness from readiness rests on this discriminant
behavior together with the within-sample separation the model exhibits, where countries at equal
readiness diverge sharply in post-quantum creation. Taiwan is absent from both indices, and a few
countries are absent from one or the other, and are excluded from the respective correlation.

This discriminant check is necessary but not sufficient, and we are explicit about what it does not
do. It shows that post-quantum control is not readiness by another name, but it is not an independent
confirmation of the post-quantum-control construct itself, because no external index measures that
construct, and the objective markers that would serve, algorithm authorship and standards
participation, are the same markers that inform the coding, so comparing against them would be
circular. A fully independent external criterion for post-quantum control, such as an audited registry
of national quantum-safe production, does not yet exist, and building or obtaining one is the single
most valuable next step for validating the central construct rather than only its separation from
readiness.

The sample is purposive rather than a census, and its inferential role should be read at three
levels. It supports the model's internal structure and discriminant behavior, since the post-quantum
axis separates from readiness within the sample and the class structure survives the sensitivity
analyses. It characterizes regional structure, since the per-continent means of Table~\ref{tab:regions}
describe a gradient across the assessed countries, read only for the continents with enough countries to
average. And it does not support a definitive global ranking of individual countries, with boundary
countries in particular reported as boundary cases rather than ranked. Because the sample is purposive
and evidence availability is associated with institutional capacity, the assessed regional picture may
overrepresent the better-documented countries, and the direction and magnitude of the bias from
countries excluded for want of public evidence cannot be established, since those countries are not
scored; they are named as a no-evidence category rather than hidden. Finally, the data are a snapshot, and fast-moving
national mandates will move some boundary cells over time, which is why the worksheet is released for
periodic re-coding rather than presented as a fixed ranking.

\section{Recommendations}\label{sec:recommend}
The findings suggest five directions for dependent countries, and for the Global South in
particular, where the readiness and sovereignty deficits are largest. The first four are
domestic, and the fifth concerns the international support without which the others may be out of
reach for the least-resourced countries. These are policy directions drawn from the assessment rather
than results derived from it, and where they compare costs or timing they state expectations that the
model does not itself quantify.

\subsection{Continental coordination}
The single most consequential gap in the regional core is the absence of any coordinating body
that sets a migration clock and a cryptographic inventory obligation, which leaves most countries
without even a starting point and without the shared urgency that a common deadline creates. A
continental mechanism, plausibly built on an existing regional institution rather than a new
one, could publish a staged timeline, require a cryptographic inventory in critical sectors, and
maintain a common register of quantum-vulnerable systems. Its value is not only technical but
economic, since a shared inventory requirement written into public procurement would aggregate
the demand of many small markets into a signal large enough to attract regional vendors, which
no single ministry can generate alone. Coordination is also, we expect, the lowest-cost of the
recommendations, because it requires convening and rule-making rather than industrial investment.

\subsection{Governance sovereignty first}
Because we expect governance sovereignty to be among the lower-cost layers to build and it is the one on which the
region already has a foothold, it is the natural first target. Several countries run national
certification authorities and issue national profiles, and a handful maintain active research
groups, yet almost none contributes to the international standards committees where the primitives
and their revisions are decided. Moving from observer to active contributor in those committees
would give a country a voice in the decisions it currently only inherits, at a cost measured in
expert time rather than capital. Governance sovereignty does not by itself remove dependence at
the algorithm and hardware layers, but it converts a country from a silent taker into a participant
that can at least contest a bad decision and prepare for the next revision.

\subsection{Sovereignty-aware procurement}
Procurement is where dependence is quietly locked in, because a national public-key
infrastructure delivered and operated by a foreign vendor is easily recorded as a sovereign
asset when it is not. Procurement rules should therefore separate the layers, asking not only
whether a capability is national in name but whether the algorithm, the vendor, the hardware, and
the key material are domestically controlled, and weighting bids accordingly. The external-dependency
construct gives such rules a concrete and auditable criterion, and applying it at the
point of purchase is expected to be far cheaper than retrofitting sovereignty after a foreign stack is
entrenched for a generation, a policy expectation rather than a cost the model quantifies.

\subsection{Indigenous research as a lever}
The region's stranded research capacity is its most direct path to genuine sovereignty, and it
is being wasted. The authored work already present, such as the Senegalese code-based submission
to an international competition and the Ethiopian administration mandated to produce cryptographic
products, shows that indigenous capability exists but is disconnected from any national migration.
Connecting these efforts to national programs, through funded mandates, procurement preferences
for domestically analyzed or authored schemes, and sustained support for the university groups
that host the talent, would convert academic output into national capacity. No dependent country
will match the largest makers, but the capable but post-quantum-dependent cases demonstrate that even
a modest indigenous base changes a country's position from pure dependence toward partial control,
which is the realistic ambition for the coming decade.

\subsection{Support and burden sharing}
The deadlines that make the transition urgent apply equally to countries that cannot fund an
industrial response, and leaving them behind is a shared risk rather than a local one, because
insecure links propagate through the interconnected systems that everyone uses. There is
therefore a case for international support that is specific rather than rhetorical, including
risk audits and inventory assistance for countries that lack the capacity to perform them, shared
access to open and independently analyzed post-quantum implementations so that a dependent country
is not forced to choose between an unvetted domestic option and a foreign proprietary one, and
training programs that build the regional talent base the other recommendations rely on. Such
support is most effective when it strengthens indigenous capacity rather than substituting for
it, because assistance that arrives as another foreign-supplied stack reproduces the dependence
it was meant to relieve. The external-dependency construct can serve donors and recipients alike
as a criterion for distinguishing support that builds sovereignty from support that entrenches
dependence.

\section{Conclusion}\label{sec:conclusion}
We presented RSCM, a national capability model that measures post-quantum readiness alongside a
decomposition of cryptographic sovereignty into indigenous capacity, post-quantum control, and
external dependency, with a maker gate that recognizes demonstrated indigenous creation in at least
one core post-quantum layer on an institutionally sustained basis. Applied to fifty-seven countries, it
shows that post-quantum creation is concentrated among makers that are almost all wealthy or
long-industrialized, that many capable countries hold strong general cryptography without having built
post-quantum control, and that the countries of Africa and the Arab world combine low readiness with
post-quantum layers resting on foreign foundations, while a research institute or a sustained national
program can carry even a middle-income country through the gate.

The decomposition is the central correction, keeping classical capability and governance authority from
being read as post-quantum sovereignty. A country that measures only how far its migration has
progressed can mistake speed for autonomy and emerge more dependent than it entered, and the evidence
that post-quantum control is reachable by sustained institution-building, yet absent in several capable
countries whose indigenous capacity is unconnected to the post-quantum layer, is what turns the
diagnosis into an agenda. Two extensions matter most. The sample can be grown toward a full census,
particularly of the low-capacity countries omitted here for lack of public evidence, and because the
data are a snapshot of a fast-moving field, a longitudinal track would show whether the transition
entrenches the current concentration of post-quantum control or begins to redistribute it. By measuring
sovereignty rather than only readiness, the model makes visible a structure that one-dimensional
assessments conceal and offers the assessed countries a concrete basis for deciding how much of their
cryptographic future they intend to control.

\section*{Data Availability}
All project files, code, and the assessment dataset are at
\url{https://github.com/drbouke/rscm-explorer}. An interactive explorer for previewing the tool and
following updates is at \url{https://drbouke.github.io/rscm-explorer/}.

\end{document}